\documentclass[12pt, onehalfspacing, leqno]{article} % use larger type; default would be 10pt
\usepackage[T1]{fontenc}

\usepackage{libertine}

\usepackage{geometry} % to change the page dimensions
\usepackage[svgnames]{xcolor}
\usepackage{graphicx} % support the \includegraphics command and options
\usepackage{tikz} % Allows for tikz support; image generation
\usetikzlibrary{trees}
\usepackage{tkz-euclide}

\usepackage{multicol} % creates mini columns within a normal page
\usetikzlibrary{hobby}

\usepackage{booktabs} % for much better looking tables
\usepackage{array} % for better arrays (eg matrices) in maths
\usepackage{paralist} % very flexible & customisable lists (eg. enumerate/itemize, etc.)
\usepackage{verbatim} % adds environment for commenting out blocks of text & for better verbatim
\usepackage{subfig} % make it possible to include more than one captioned figure/table in a single float
\usepackage{amsmath, amssymb, amsthm, thmtools, euscript} %basic math symbols, scripts, theorem environments
\usepackage{mathrsfs}
\usepackage{mhsetup, mathtools} %Must be used together; 
\newtheorem{proposition}{Proposition}  %[section] % To number by section 0.1, 1.2, etc
\newtheorem{lemma}{Lemma}

\newtheorem{theorem}{Theorem}
\newtheorem{corollary}{Corollary}

\theoremstyle{definition}
\newtheorem{definition}{Definition}
\newtheorem{axiom}{Axiom}
\newtheorem{example}{Example}
\newtheorem{remark}{Remark}

\newcommand{\R}{\mathbb{R}}

\newcommand{\F}{\mathscr{F}}

\newcommand{\ra}{\rightarrow}

\newcommand{\RA}{\Rightarrow}
\newcommand{\LRA}{\Leftrightarrow}

\definecolor{purple}{RGB}{85, 6,139}
\definecolor{teal}{RGB}{2,108,128}
\definecolor{lavender}{RGB}{129, 102, 122}
\definecolor{carolina blue}{RGB}{68, 157, 209}
\definecolor{phthalo blue}{RGB}{2, 8, 135}
\definecolor{purple2}{RGB}{149, 96, 219}
\definecolor{green1}{RGB}{96, 219, 117}

\usepackage{natbib}
\usepackage[colorlinks]{hyperref}

\usepackage{hyperref}
\usepackage{nameref}
\hypersetup{
breaklinks=true,
  colorlinks   = true, %Color links instead of ugly boxes
  urlcolor     = purple, %Colour for external hyperlinks
  linkcolor    = teal, %Colour of internal links
  citecolor   = gray%gray%red!40!black} %Colour of citations
}
\usepackage{fancyhdr} % This should be set AFTER setting up the page geometry
\usepackage{sectsty}
\usepackage{setspace}
\allsectionsfont{\mdseries\upshape} % (See the fntguide.pdf for font help)
\usepackage[nottoc,notlof,notlot]{tocbibind} % Put the bibliography in the ToC
\usepackage[titles,subfigure]{tocloft} % Alter the style of the Table of Contents

\title{\textsc{Ambiguity and Partial Bayesian Updating}\thanks{I would like to thank Pietro Ortoleva and Federico Echenique for their guidance and support through all stages of this project.  I would also like to thank Adam Dominiak, Kota Saito, Euncheol Shin, Hector Tzavellas, and Gerelt Tserenjigmid.  Much of this work is derived from my second-year paper in the Social Science PhD program at Caltech, and was also the second chapter of my dissertation. All errors are my own.}  } %%%   TITLE %%%
\author{Matthew Kovach\footnote{Department of Economics, Virginia Tech.  E-mail: mkovach@vt.edu}}
\date{First Version: November 22, 2015; Current Version: March 15, 2023}

\begin{document}
\maketitle

\vspace{7 mm}

\noindent{\textbf{Abstract:} 
Models of updating a set of priors either do not allow a decision maker to make inference about her priors (full bayesian updating or FB) or require an extreme degree of selection (maximum likelihood updating or ML). I characterize a general method for updating a set of priors, \textbf{partial bayesian updating} (PB), in which the decision maker (i) utilizes an event-dependent threshold to determine whether a prior is likely enough, conditional on observed information, and then (ii) applies Bayes' rule to the sufficiently likely priors. I show that PB nests FB and ML and explore its behavioral properties.

\vspace{7 mm}

\noindent{\textbf{Keywords:} Ambiguity Aversion, Dynamic Consistency, Full Bayesian Updating, Maximum Likelihood Updating, Partial Bayesian Updating

\vspace{30 mm}

\pagebreak

\section{Introduction}

Since \citet{ellsberg1961} noted a distinction between ambiguity and risk, numerous models of ambiguity averse agents have been proposed.  One of the earliest and most well-known models of ambiguity aversion is the Maxmin Expected Utility (MEU) model (\citet{Gilboa1989}). Under MEU, an agent entertains multiple possible beliefs, or a set of priors (represented by a closed, convex set of probability distributions), and she evaluates an act by its worst-case expected utility. However, extending MEU preferences to dynamic contexts poses a challenge. Unlike models in which beliefs are given by a single probability (such as subjective expected utility, or SEU), where the natural notion of updating is Bayes' rule, there is no obvious way in which a set of beliefs should be updated.

To illustrate, suppose Kate is trying to save for her retirement.  She is concerned that she may not know the true distribution of outcomes, and so she decides to consult a panel of experts. Each expert gives her a different picture of the economy, or a prior.\footnote{In this paper, the set of priors is purely subjective. However, the analogy of a prior as a possible model is useful for illustration.}     Kate's world is quite simple: the economy can grow ($g$), remain constant ($c$), or shrink ($s$).  The states of the world are $S=\{g, c, s\}$, and each expert has provided Kate with a probability distribution over these three states.\footnote{Ambiguity models explicitly assume the decision maker does not reduce all the priors into a single prior, otherwise the she would be indistinguishable from a standard Bayesian.} For simplicity, suppose Kate determines her initial set of beliefs from the distributions provided by these experts: $\pi_1=(\frac{7}{12}, \frac13, \frac{1}{12}), \pi_2=(\frac{1}{6}, \frac23, \frac{1}{6})$ and $\pi_3=(\frac{1}{12}, \frac13, \frac{7}{12})$.\footnote{Under maxmin expected utility (MEU) preferences \citep{Gilboa1989}, the set of priors $\{\pi_1, \pi_2, \pi_3\}$ and its closed, convex hull are behaviorally equivalent, and so we say she has the set of priors $\mathscr{C}=conv(\{\pi_1, \pi_2, \pi_3\})$. Conceptually, it is as if Kate takes the expert forecasts as ``bounds'' on the truth.} 
Suppose Kate learns that there has been an ``increase in unemployment,'' which rules out $g$; the true state belongs to $\{c,s\}$.  How should Kate incorporate this new information into her beliefs? Kate can use the information to ``update" each of the experts' forecasts, but she can also use the information to discriminate between the experts. Kate's problem is the topic of this paper.

There are two well-known procedures for updating a set of priors explored in the literature. In the first procedure, known as full bayesian updating (FB) \citep{gilboa2011}, the agent applies Bayes’ rule to each prior.  In the second procedure, known as maximum likelihood updating (ML) \citep{gilboa2011}, the agent only retains priors that assigned the greatest probability to the observed event.  In the case of Kate, if she practices FB she continues to consider $\pi_1$ for her decision making, even after observing the increase in unemployment.  If she practices ML, then she only considers $\pi_3$.

Both methods of updating a set of priors may be unsatisfactory. Full bayesian updating requires one to treat all priors as ``equally plausible,'' even in the face of evidence that is highly unlikely under certain priors. That is, she cannot use the new information to make an inference about which priors to believe.  Maximum likelihood updating does not exhibit this problem, but it requires an ``extreme'' response to information.  Returning to Kate, under FB she treats experts who were good predictors the same as those who were very bad, continuing to put equal stock in $\pi_1$ and $\pi_3$. It is reasonable that she might lose confidence in the expert that suggested $\pi_1$ and wish to ignore his advice going forward. In contrast, under ML she only believes those experts that gave the highest probability to the outcome observed.  Kate puts all her faith in $\pi_3$, even though under $\pi_2$ the observed information (e.g., increase in unemployment) is quite likely ($\pi_2(\{c,s\})=\frac56$). Kate might find it imprudent to ignore advice that was reasonably predictive of what happened. 

In this paper, I propose an alternative updating procedure that generalizes the above two procedures: \textbf{partial bayesian updating (PB)}. Under PB, after observing some event $A$, the decision maker determines a selection of sufficiently likely priors and then applies Bayes' rule to these priors.  %Formally, given an ex-ante set of priors $\mathscr{C}$ and a non-null event $A \subset S$, the PB posterior set is 
%\begin{equation}\label{updating} \mathscr{C}_A^{\rho}= \{\pi_A |  \pi(A) \geq \rho(A) \max_{\mu \in \mathscr{C}}\mu(A)  \text{ and }  \pi \in \mathscr{C}\},\end{equation}
%where $\pi_A$ is the Bayesian update of $\pi$ given $A$.  
If Kate is a partial bayesian updater, she uses the observation of an event to infer which priors are still plausible enough. A prior is deemed sufficiently plausible if it passes a likelihood ratio test with an information dependent threshold taking values between zero and one. I refer to this as her \textbf{inference threshold} after $A$.  Whenever a prior belief and the observed event ``conflict,'' that prior is deemed insufficiently plausible and is discarded. The retained priors are then updated according to Bayes' rule.  I formally define and further discuss the model in \autoref{model}. 

%$\rho(A) \in [0,1]$. I refer to $\rho(A)$ as her \textbf{inference threshold} after $A$.  In other words, whenever a prior belief and the observed event ``conflict,'' $\pi(A) <  \rho(A) \max_{\mu \in \mathscr{C}}\mu(A)$, that prior is deemed not sufficiently plausible and is discarded. The retained priors are then updated according to Bayes' rule.  

Behaviorally, the inference threshold represents how stringent her test is for the retention of prior beliefs. When the threshold is sufficiently low, PB coincides with FB; everything passes and she makes no inference. When the threshold is one, PB coincides with ML; only maximally predictive models pass and she makes ``maximal'' inference.  For intermediate values, PB can be viewed as a compromise between an impulse to be ``cautious'' (FB) and a desire to ``reach a conclusion'' (ML).  Importantly, since the inference threshold may vary from event to event, Kate's reaction to information may vary depending on her subjective ability to discriminate between priors \emph{at a particular event}. For instance, if an information source is perceived as less reliable or precise, she may be reluctant to make an inference, captured by a lower threshold.  Consequently, her propensity to exhibit dynamic reversals depends one her perception of the event.

%
%Behaviorally, the inference threshold $\rho(A)$ represents how stringent her test is for the retention of prior beliefs. When $\rho(A)=0$, PB coincides with FB; everything passes and she makes no inference. For $\rho(A)=1$, PB coincides with ML; only maximally predictive models pass and she makes ``maximal'' inference.  For intermediate values, PB can be viewed as a compromise between an impulse to be ``cautious'' (FB) and a desire to ``reach a conclusion'' (ML).  Importantly, since $\rho$ may vary from event to event, Kate's reaction to information may vary depending on her subjective ability to discriminate between priors \emph{at a particular event}. For instance, if an information source is perceived as less reliable or precise, she may be reluctant to make an inference (captured by a lower $\rho(A)$).  Consequently, her propensity to exhibit dynamic reversals depends one her perception of the event.  

While partial bayesian updating always results in a posterior set that is between full bayesian and maximum likelihood, it can result in fundamentally different behavior. Indeed, it may result in choice behavior that is distinct from either model. Recalling Kate, there may exist investments $f$ and $g$ such that $f$ is preferred to $g$ under both FB and ML, but under PB she strictly prefers $g$ to $f$ for some range of inference thresholds.  %$\rho(A)$. 

%Formally, it is easy to see that for every $A$ the posterior sets are nested: $\mathscr{C}^{ML}_A \subseteq \mathscr{C}^{\rho}_A \subseteq \mathscr{C}^{FB}_A$. However, there may exist acts $f,g$ and event $A$ where $f$ is preferred to $g$ under both FB and ML, $f \succsim_{FB(A)}g$ and  $f\succsim_{ML(A)}g$, but $g \succ_{\rho(A)} f$, where the preference relations $\succsim_{FB(A)}, \succsim_{ML(A)}$, and $\succsim_{\rho(A)}$ denote the conditional preference derived from a common \emph{ex-ante} preference $\succsim$ according to FB, ML, and PB, respectively (see \autoref{reversal}). 

%(\autoref{updating}) 
I provide behavioral foundations for PB with two novel axioms (and a collection of standard postulates) in \autoref{characterization}. The first axiom, \nameref{WUDC}, is a weak notion of dynamic consistency that ensures a form of planning consistency when comparing acts to ``sure things.'' 

Returning to Kate's problem, suppose there are two investments, one that has uncertainty, $f$, and one that is safe, $x$.   If Kate would prefer exposure to $f$ only in the ``increased unemployment'' event,  $\{c,s\}$, and $x$ otherwise, then after learning that the economy is not growing she must prefer $f$ to $x$. The second novel axiom, \nameref{URC}, requires that her inferences are monotonic in the probability that her priors assign to the observed information. If Kate decides to keep $\pi_2$, she must also keep $\pi_3$ because the observed event, $\{c,s\}$, is more likely under $\pi_3$ than $\pi_2$. These two axioms, combined with postulates ensuring an MEU representation, are equivalent to PB.%: the agent's conditional beliefs are determined according to \autoref{updating}.

I provide further analysis of PB in \autoref{discussion}. First, I discuss comparative statics for the inference parameter $\rho$, and thereby establish a connection between conditional ambiguity attitudes and comparative inference under PB.  Under PB, greater inference (larger $\rho(A)$) is equivalent to being less ambiguity averse after $A$. A simple corollary to this finding is that the various updating rules are ordered by how ambiguity averse the corresponding conditional preferences are: (i) FB results in the most ambiguity averse conditional preference; (ii) ML results in the least ambiguity averse; and (iii) PB results in something in between. 

Second, I show that PB may result in a \emph{primacy effect}: initial information may have a disproportionate impact on later beliefs due to early inferences.  The sensitivity of beliefs to the order in which information is received is well-known in the psychology literature (See \cite{hogarth1992} for a discussion).  To show this, I provide a formal definition of Informational Path Independence and construct a set of priors such that PB violates it. Consequently, ML also exhibits a primacy effect.  In fact, the only special case of PB that is sure to satisfy path independence is FB ($\rho=0$). 

Lastly, I characterize the special case of PB where the inference parameter is event independent, $\rho_A=\rho$ for every $A$, which was utilized to study learning in \citet{Epstein2007} and asset pricing in \citet{Epstein2008ambiguity}. The axiomatization is achieved by strengthening the monotonicity condition to a form of relative monotonicity across events: \nameref{ORC}.  To understand \nameref{ORC}, consider Kate and suppose that after learning $A=\{s, c\}$ (i.e., unemployment has increased) she retains $\pi_2$. What should her posterior beliefs be if she had instead learned $B=\{g, c\}$ (i.e., unemployment has decreased)? Since both $A$ and $B$ are equally likely according to $\pi_2$ ($A$ and $B$ are equally "informative" about $\pi_2$), then we might expect Kate to also retain $\pi_2$ after $B$. \nameref{ORC} strengthens \nameref{URC} to impose this form of consistency across events.

\subsection{Related literature}

\cite{Gilboa1993} axiomatized ML updating when the set of priors is determined by a convex capacity, hence preferences admit both MEU and Choquet expected utility (CEU) representations. This characterization of ML however does not extend to general MEU preferences, which I work with. Full bayesian updating was first proposed by \citet{jaffray1988} and was later axiomatized by \citet{pires2002} through the use of ``conditional certainty equivalents."  It was later shown by \citet{Ghirardato2008} that FB is equivalent to dynamic consistency of the agent's unambiguous preference. Recently, \cite{Hill2021} develops a model of updating in which a decision maker utilizes a confidence ranking to update some beliefs. This model includes PB as a special case, and consequently it also includes FB and ML. However, the specific characterization of PB is not addressed by \cite{Hill2021} and our axiomatizations are different. \cite{Cheng2021} develops the ``Relative Maximum Likelihood'' (RML) rule which also generalizes FB and ML. In the RML, the agent linearly contracts her set of priors towards the maximum likelihood priors and then applies Bayes' rule. This rule violates the monotonic selection of priors (\nameref{URC}) and so is generally distinct from PB (see section \ref{rml} for details). 

In general, it is difficult to maintain dynamic consistency and consequentialism in the MEU model (see \cite{gilboa2011} for an excellent overview). Under certain restrictions, the MEU model may sometimes maintain dynamic consistency and consequentialism. \citet{Epstein2003} show that dynamic consistency and consequentialism may be maintained in a multiple priors model with a restriction on the information structure (a filtration) and a rectangularity condition (with respect to the fixed filtration) on the set of priors $\mathscr{C}$. Rectangularity is therefore a joint restriction on the evolution of uncertainty and the set of priors.  Under rectangularity, FB, ML and PB are all equivalent. Since PB relaxes dynamic consistency it does not require rectangularity and so it may be applied to any set of priors and does not restrict the information structure. Relatedly, \cite{dominiak2011} show that dynamic consistency for CEU preferences is possible when information is represented by a filtration and the terminal events are unambiguous.  \citet{Hanany2007} \citep{Hanany2009} show that Dynamic Consistency for multiple prior preferences (ambiguity averse preferences) requires a form of "history dependent" updating, so that preferences are dependent on the feasible set and past choices. In contrast, PB only depends on the realized information.  

In applications, FB updating has received the most attention. For instance, \citet{Beauchene2019} study the effect of ambiguity in persuasion,  \citet{Kellner2018} study ambiguity in cheap talk games, while \citet{Bose2014} study the effect of ambiguous communication in mechanism design. All three utilize MEU and FB updating. 

However, rules similar to PB have been applied by \citet{Epstein2007} and \citet{Epstein2008ambiguity} to study long-run learning under ambiguity and asset pricing, respectively.  They do not provide an axiomatization of this type of updating, as their focus is on studying when ambiguity can be resolved over time and on how asset prices react to news. Further, they study the case of event-independent inference threshold: $\rho(A)=\rho$ for every $A$.  In contrast, this paper focuses on the behavioral foundations of PB updating and allows for an event-dependent inference threshold: $\rho(A)$ may vary across events.

\section{Model and notation}\label{model}

I utilize a standard setting for decision making under uncertainty. There is a finite set of states of the world, $S$ and an algebra of events $\Sigma \subseteq 2^S$. The set of consequences, $X$, is assumed to be a convex subset of a vector space.\footnote{This assumption also appears in \cite{Ghirardato2004}. Notice that when $X$ is the set of lotteries over some prize space, this   coincides with the setting of \citet{Anscombe1963}.} An act is a function $f:S \rightarrow X$, and $\mathscr{F}$ denotes the set of all acts.  Following a standard abuse of notation, for any $x \in X$, I mean by $x \in \mathscr{F}$ the constant act that returns $x$ in every state.  For any $f, g \in \mathscr{F}$ and for any $A \in \Sigma$, let $fAg$ denote the act that returns $f(s)$ when $s \in A$ and returns $g(s)$ when $s \in A^c \equiv S\backslash A$.  Since $X$ is convex, mixed acts can be defined pointwise so that for every $f, g \in \mathscr{F}$ and $\lambda \in [0,1]$, by $\lambda f + (1-\lambda)g$ I mean the act that returns $\lambda f(s) + (1-\lambda)g(s)$ for each $s \in S$.  Finally, call $A$ \emph{unambiguously $\succsim$-nonnull} if for all $x,y \in X$ such that $x\succ y$, it follows that $xAy \succ y$. %Under MEU preferences this ensures that $\pi(A) > 0$ for all $\pi \in \mathscr{C}$. 

In this paper I take the conditional preference approach. Formally, the agent has a collection of preference relations, $\{\succsim_A \}_{A \in \Sigma}$ over $\mathscr{F}$.  For each $A \in \Sigma$, $\succ_A$ and $\sim_A$ represent the asymmetric and symmetric parts of $\succsim_A$.  When the only information the agent has is the entire state space (i.e., the agent has no information), I refer to this relation as the unconditional preference relation.  The interpretation is that after observing some event $A$, the agent updates her preferences from $\succsim :=\succsim_{S}$ to $\succsim_{A}$.  
%Given a utility  $u:X \ra \R$ and a closed, convex set $\mathscr{C} \subset \Delta(S)$, we define the MEU functional as
%\[ U(f)= \min_{\pi \in \mathscr{C}}\mathbb{E}_{\pi}[u(f)].\]
\begin{definition}\label{MEUdef} A preference relation $\succsim$ admits a \textbf{Maxmin Expected Utility} (MEU) representation if and only if there exists a closed, convex set of priors $\mathscr{C}$ and a non-constant affine function $u:X \rightarrow \mathbb{R}$ so that \[f \succsim g \LRA \min_{\pi \in \mathscr{C}} \sum_{s \in S}u(f(s))\pi(s) \geq \min_{\pi \in \mathscr{C}} \sum_{s \in S}u(g(s))\pi(s).\]  
%\item[(ii)] If $x \succsimthe collection additionally satisfies ordinal preference consistency, we can suppose without loss that for all $A$, $u_A = u_{S}$.  
\end{definition}

Throughout the paper, I will utilize the following facts and notation.  
\begin{itemize}
\item Each conditional preference $ \succsim_A $ admits an MEU representation $(u_A, \mathscr{C}_A)$ as in \autoref{MEUdef} (this will follow from \autoref{CMEU}).
\item When the collection of preferences $\{\succsim_A \}_{A \in \Sigma}$ satisfies ordinal preference consistency (see \autoref{CMEU}), we can suppose without loss that for all $A$, $u_A = u_{S}$.  
\item If $A$ is unambiguously $\succsim$-nonnull, then $\pi(A)>0$ for all $\pi \in \mathscr{C}$.
\item To simplify notation throughout, I will write 
\[ U(f):= \min_{\pi \in \mathscr{C}}\sum_{s \in S}u(f(s))\pi(s).\] 
\end{itemize}

%To simplify notation throughout, I will write 
%\[ U(f):= \min_{\pi \in \mathscr{C}}\sum_{s \in S}u(f(s))\pi(s), \] %and 
%%\[ U_A(f):= \min_{\pi \in \mathscr{C}_A}\sum_{s \in S}u(f(s))\pi(s) .\]
%

For any $\pi \in \Delta(S)$, and $A\subseteq S$ with $\pi(A)>0$, let $\pi_A$ denote the Bayesian update of $\pi$ conditional on $A$: $\pi_A(s)= \frac{\pi(s)}{\pi(A)}$ for $s \in A$, $0$ otherwise. Let $[\pi_A]=\{\hat{\pi} \in \Delta(S) \mid \hat{\pi}_A=\pi_A\}$, which is the set of distributions that result in the same posterior under Bayes' rule. For any $\Pi \subseteq \Delta(S)$ and $A$ such that $ \pi (A) >0$ for all $\pi \in \Pi$, $BU(\Pi, A)=\{ \pi_A \mid \pi \in \Pi\}$ is the set of prior-by-prior updates. 

Given any $\mathscr{C} \subseteq  \Delta(S)$ and an event $A$ with $\pi(A)>0$ for all $\pi \in \mathscr{C}$, 
\begin{itemize}
\item The \textbf{full bayesian} (FB) update of $\mathscr{C}$ given $A$ is: \[\mathscr{C}_A^{FB}:=BU(\mathscr{C}, A).\]
\item  The \textbf{maximum likelihood} (ML) update of $\mathscr{C}$ given $A$ is:  \[\mathscr{C}_A^{ML}:= BU(\{\pi \mid \pi(A)= \max_{\mu \in \mathscr{C}}\mu(A) \text{ and } \pi \in \mathscr{C}\}, A).\] 
\end{itemize}

%Further, for any $A$ for which \[ U_A(f)= \min_{\pi \in \mathscr{C}}\mathbb{E}_{\pi}[u(f)], \]

%\begin{definition}A preference relation $\succsim$ admits a Maxmin Expected Utility (MEU) representation if there if there exist a $u:X \ra \R$, a closed, convex set $\mathscr{C} \subset \Delta(S)$, such that \[f  \succsim g \text{ if and only if } \mathbb{E}_\pi
%
%\end{definition}

\begin{definition}Say that a collection of preferences $\{\succsim_A\}_{A\in\Sigma}$ admits a \textbf{partial bayesian updating} (PB) representation if there exist $u:X \ra \R$, a closed, convex set $\mathscr{C} \subset \Delta(S)$, and for each unambiguously $\succsim$-nonnull $A \in \Sigma$ there exists a $\rho(A) \in [0,1]$ such that 
\begin{itemize}
\item[(i)] $\succsim$ admits a MEU representation $(u,\mathscr{C}),$
\item[(ii)] $\succsim_A $ admits a MEU representation $(u,\mathscr{C}_A^{\rho}),$ where\begin{equation*}\mathscr{C}_A^{\rho} = \{\pi_A \mid  \pi(A) \geq \rho(A) \max_{\mu \in \mathscr{C}}\mu(A)  \text{ and }  \pi \in \mathscr{C}\}.\end{equation*}
\end{itemize}
If there is a $\rho \in [0,1]$ such that $\rho(A)=\rho$ for every $A$, say that preferences admit a constant PB representation. For simplicity, I will offer refer to a PB represetnation as a triple $(u, \mathscr{C}, \rho)$. 
\end{definition}

A partial bayesian updater uses the observation of an event to infer which priors are still likely enough. The notion of likely enough corresponds to a likelihood ratio test, where a prior passes when $\pi(A) \geq \rho(A)\max_{\mu \in \mathscr{C}}\mu(A)$, or \[\frac{\pi(A)}{\max_{\mu \in \mathscr{C}}\mu(A)} \geq \rho(A).\] 
The priors that pass are then updated according to Bayes' rule. The threshold, $\rho(A)$, is event-dependent and subjective. I will refer to $\rho$ as the agent's \textbf{inference threshold}. 

When $\rho(A)= 0$, every prior passes the test and the agent makes no inference; she engages in full bayesian updating (at $A$). When $\rho(A)=1$, the agent makes maximal inference and updates only those priors that gave the greatest likelihood to $A$; she engages in maximum likelihood updating.  Therefore PB nests FB and ML as special cases. Further, since $\rho$ is event dependent, a decision maker might appear to be closer to FB after some events while appearing closer to ML after other events. 

To facilitate comparisons between PB updating and both FB and ML, I will sometimes make use of the following notation. Given a preference $\succsim$ and an event $A$, 
\begin{enumerate}
\item $U_A^{\rho}$  denotes the conditional utility under PB with inference threshold  $\rho$.  
\item The conditional preference under FB updating is denoted $\succsim_{FB(A)}$ with corresponding utility $U_A^{FB}$. 
\item The conditional preference under ML updating is denoted $\succsim_{ML(A)}$ with corresponding utility $U_A^{ML}$. 
\end{enumerate}

%
%\begin{example} Let $p^* \in \Delta(S)$ and let $\mathscr{C}=\{p \mid d(p,p^*) \leq r \}.$ When $d$ is relative entropy, we have the common specification of beliefs in the robust control literature \citep{hansen2001}. As noted in \cite{Epstein2003}, this set is non-rectangular.
%\end{example}

\begin{example}\label{reversal} Kate is presented with a slight variation of the standard Ellsberg urn, which contains some composition of red, blue, and yellow balls. She will be paid according the color of the drawn ball, hence there are three states of the world, $S =\{r,b,y\}$.  The true distribution over the states of the the world belongs to 
\[\mathscr{C} = conv\left\{ \left(\frac{9}{20}, \frac{9}{20}, \frac{2}{20}\right), \left(\frac{6}{10}, \frac{2}{10}, \frac{2}{10}\right), \left(\frac{1}{10}, \frac{3}{10}, \frac{6}{10}\right) \right\}.\] For ease of reference, let $\pi_1=(\frac{9}{20}, \frac{9}{20}, \frac{2}{20}), \pi_2=(\frac{6}{10}, \frac{2}{10}, \frac{2}{10}), \pi_3=(\frac{1}{10}, \frac{3}{10}, \frac{6}{10})$.

I will illustrate Kate's problem with the follow two acts. 

\begin{table}[h]
\centering
\begin{tabular}{ l c c c }
 &$r$ &$b$ &$y$  \\
\hline 
 $f$ & $0$ & $10$ & $0$ \\
 $g$ & $10$ & $0$ & $0$ \\
\end{tabular}
\end{table}

Before any information, it is straightforward to calculate that $U(f)=2$ and $U(g)=1$. Suppose Kate is provided with interim information that the drawn ball is not yellow: the true state is in $A = \{r,b\}$.\footnote{While I allow for any event, restricting the set of events to tree-structure as in \citet{Epstein2003} would not substantively change things. To see this, note that $\mathscr{C}$ is not rectangular with respect to $\{\{r,b\},\{y\}\}$, meaning that FB and ML result in different posterior beliefs and also fundamentally distinct behavior.} How does her ranking change?

Notice that $\pi_1(A) > \pi_2(A) > \frac12 > \pi_3(A)$. According to both $\pi_1$ and $\pi_2$, $A$ is ``likely'' to occur, and so it seems reasonable that Kate continues to believe they are plausible after $A$. In contrast, $A$ is ``unlikely'' to occur according to $\pi_3$, and so Kate might infer that $\pi_3$ is implausible, but only after $A$ has in fact happened.\footnote{Indeed, she would presumably deem $\pi_3$ very relevant if she had instead learned that the state was either blue or yellow: $\{b,y\}$.} Consequently, $g \succ_A f$ seems reasonable. This is a reversal of her ex-ante ranking $f \succ g$, and so she violates dynamic consistency. But she does so because of her reasonable inference.
 
Put another way, Kate knows two facts: (i) the probability of $b$ is relatively high in comparison to $r$ when the probability of $y$ is high, and (ii) $y$ did not occur. Given these facts, she infers that \emph{$r$ is relatively more likely than $b$}; hence her posterior set shrinks the probability of $b$ relative to $r$.

\begin{figure}[h]
\centering
\includegraphics[height=7cm]{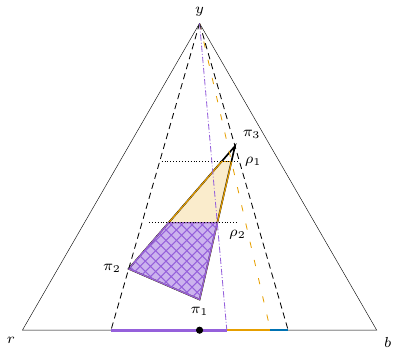}
\caption{$\mathscr{C}$ and posterior beliefs after $A=\{r,b\}$.}\label{reversalfig}
\end{figure}

As $f$ and $g$ are symmetric on $A$, if Kate updates her beliefs with either FB or ML, she must be indifferent between these acts: $f \sim_{FB(A)}g,$ and $f \sim_{ML(A)}g$.  This can be easily seen by directly calculating the posterior sets or the corresponding utilities (see \autoref{Ex1-utility}). Under FB, $\mathscr{C}^{FB}_A = conv\left\{ \left(\frac{3}{4}, \frac{1}{4}\right), \left(\frac{1}{4}, \frac{3}{4} \right)\right\},$ which is the entire region between the black, dashed lines in \autoref{reversalfig} (the union of the purple, orange and blue sections of the base). Under ML, $\mathscr{C}^{ML}_A = \left\{ \left(\frac{1}{2}, \frac{1}{2}\right) \right\},$ which is the black circle at the base of \autoref{reversalfig}.

To illustrate PB, consider two values of the inference threshold, $\rho_1=\frac12$ and $\rho_2=\frac{13}{18}\approx 0.72$.\footnote{Note that whenever $\rho \in (\frac49, 1)$, the posterior beliefs will be strictly between FB and ML. For $\rho \leq \frac49$, PB coincides with FB, while for $\rho=1$, PB reduces to ML.} Then
\[\mathscr{C}^{\rho_1} = \left\{ \left(\frac{3}{4}, \frac{1}{4}\right), \left(\frac{3}{10}, \frac{7}{10}\right)\right\} \text{ and } \mathscr{C}^{\rho_2} = \left\{ \left(\frac{3}{4}, \frac{1}{4}\right), \left(\frac{11}{26}, \frac{15}{26}\right)\right\}.\]
In \autoref{reversalfig}, $\mathscr{C}^{\rho_1}$ is the region between the left-most dashed line and the loosely dashed line (union of the purple and orange sections at the base of the simplex), which is the projection of the (purple) cross-hatched and the (orange) shaded regions. $\mathscr{C}^{\rho_2}$ is the region between the left-most dashed line and the dash-dotted line (the purple section at the base of the simplex), which is the projection of the (purple) crosshatched region. The posterior set shrinks only on the right side (the blue portion is dropped and then the orange portion), because $\pi_2$ passes the likelihood test for both example values of $\rho$.  Note that as $\rho$ increases, the posterior set shrinks. 

To further illustrate the updating rules, the resulting utilities for $f$ and $g$ under the various updating rules are shown in \autoref{Ex1-utility}.

\begin{table}[h]
\centering
\begin{tabular}{  l  c  c  c  c }
			  &  $U_A^{FB}$ & $U_A^{\rho_1}$ & $U_A^{\rho_2}$ & $U_A^{ML}$ \\ \hline \hline
			 $f$  & $2.5$   & $2.5$ & $2.5$ & $5$  \\  
			 $g$ & $2.5$  & $3$   & $4.23$ & $5$\\  
		\end{tabular}
\caption{Conditional utilities for $f$ and $g$ after $A$ under PB for $\rho_1(A)=1/2$ and $\rho_2(A)=13/18$, along with FB and ML.}\label{Ex1-utility}
\end{table}

Note that $f$ and $g$ both provide the same utility under FB and ML. Under PB, the utility for $f$ and $g$ increase towards the ML value as $\rho$ increases. Because of the ``skewed'' geometry of $\mathscr{C}$, the utility for $g$ increases for lower inference thresholds than $f$. The conditional utility for $f$ will begin to increase once $\rho(A)$ exceeds $\frac{\pi_2(A)}{\pi_1(A)}$. 

\end{example}

\begin{example} A common parametric specification in applications is the $\varepsilon$-\emph{contamination} model (see \cite{Kopylov2009, Kopylov2016} for a discussion). In this model, the set of priors is a mixture between a reference $\pi^*$ and a set of possible priors representing the agent's uncertainty: $\mathscr{C}=(1-\varepsilon)\pi^* + \varepsilon\Pi$ for $\Pi \subseteq \Delta(S)$.  As this is a special case of MEU, it is simple to directly apply PB.  In this case, the posterior set depends on both the agent's confidence in the reference prior, given by $1-\varepsilon$, and her inference threshold $\rho(A)$.  A prior $\pi \in \mathscr{C}$ is updated after event $A$ if and only if $\pi(A) \ge \rho(A)[ (1-\varepsilon)\pi^*(A) + \varepsilon \overline{m}_A(\Pi)]$, where $\overline{m}_A(\Pi):= \max_{p \in \Pi}p(A)$. The reference prior $\pi^*$ is retained if and only if \[\pi^*(A) \ge \frac{\rho(A) \varepsilon}{1-(1-\varepsilon)\rho(A)}\overline{m}_A(\Pi).\]
\end{example}

\section{Behavioral foundations}\label{characterization}

This section presents the basic model of preferences used in the paper. I begin by presenting the standard postulates. For simplicity, I combine them into a single axiom.

\begin{axiom}[Conditional MEU]\label{CMEU} For each $A \in \Sigma$, the preference relation $\succsim_A$ satisfies:
\begin{itemize}
\item[]{\bf Weak Order:} For  all $f, g, h \in \mathscr{F}$: (i) either $f\succsim_A g$ or $g \succsim_A f$  and (ii) if $f\succsim_A g$ and $g \succsim_A h$, then $f\succsim_A h$.
\item[]{\bf Certainty Independence:} For all $f, g \in \mathscr{F}, x \in X,$ and $\lambda \in (0,1]$, $f \succsim_A g \Leftrightarrow \lambda f + (1-\lambda)x \succsim_A \lambda g + (1-\lambda)x$.
\item[]{\bf Continuity:} For all $f, g, h \in \mathscr{F}$ , if $f \succ_A g$ and $g\succ_A h$, then there exist weights $\lambda, \gamma \in (0,1)$ such that $\lambda f + (1-\lambda)h \succ_A g$ and $g \succ_A \gamma f + (1-\gamma)h$
\item[]{\bf Monotonicity:} If $f, g \in \mathscr{F}$ and $f(s)\succsim_A g(s)$ for all $s \in A$, then $f \succsim_A g$.  %If in addition there is some $s \in A$ such that $f(s) \succ_A g(s)$, then $f \succ_A g$. 
\item[]{\bf Ambiguity Aversion:} For all $f, g \in \mathscr{F}$, if $f \sim_A g$ then for all $\lambda \in [0,1]$, $\lambda f + (1-\lambda)g \succsim_A f$.
\item[]{\bf Ordinal Preference Consistency:} For all $x,y \in \F$,  $x\succsim_A y $ if and only if $x \succsim y$. 
\item[]{\bf Non-degeneracy:} There are $f, g \in \mathscr{F}$ such that $f\succ g$. 
\end{itemize}
Since these conditions are imposed on every $\succsim_A$, for simplicity I will simply say that the collection $\{\succsim_A\}_{A \in \Sigma}$ satisfies \nameref{CMEU}.
\end{axiom}

\nameref{CMEU} is comprised of standard conditions known to be equivalent to the MEU representation \citep{Gilboa1989} plus ordinal preference consistency. Ordinal preference consistency is the requirement that tastes remain unchanged after information. Hence preference changes are due to changes in the agent's beliefs in response to the information. The conditions are applied to each preference in the collection $\{\succsim_A\}$, so that \emph{ex-ante} and \emph{ex-post} preferences have the same structure.

%I first state without proof some basic results which are used in all following proofs.  

\begin{axiom}[Consequentialism]\label{CON} For each $A \in \Sigma$ and for all $f, g \in \mathscr{F}$, \[f(s)=g(s) \mbox{ for all }s \in A \implies f \sim_A g.\]
Since this condition is imposed on every $\succsim_A$, for simplicity I will simply say that the collection $\{\succsim_A\}_{A \in \Sigma}$ satisfies \nameref{CON}
\end{axiom}

\nameref{CON} is a classic rationality condition imposed on preferences under uncertainty.  It requires that preferences conditional on $A$ only depend on how acts perform within $A$. In essence, this ensures that when the agent is told $A$, she believes $A$ has occurred and no longer considers states outside of $A$ payoff-relevant.

Another rationality condition commonly imposed on preferences under uncertainty is dynamic consistency. 

\begin{definition}\label{dc} Let $\{\succsim_A\}_{A \in \Sigma}$ be a collection of conditional preferences. For any $A \in \Sigma$, say that preferences satisfy \textbf{dynamic consistency at $A$} if for every $f,g \in \F$, both of the following hold
\begin{itemize}
\item[(a)] $fAg \succsim g \implies f\succsim_A g$,
\item[(b)] $f\succsim_A g \implies fAg \succsim g$.
\end{itemize}
If dynamic consistency holds for every $A \in \Sigma$ such that $A$ is unambiguously $\succsim$-nonnull, say that the collection $\{\succsim_A\}_{A \in \Sigma}$ is dynamically consistent. 
\end{definition}

As discussed in \citet{Ghirardato2002}, dynamic consistency links unconditional and conditional preferences through (i) a forward-looking planning condition and (ii) a backward-looking condition ensuring that an agent who prefers $f$ to $g$ after observing $A$ would be willing to ``commit'' to the plan $f$ if $A$ and $g$ if $A^c$.\footnote{The version of dynamic consistency utilized here is from \citet{Ghirardato2002}, which differs slightly from the form appearing in \citet{Epstein2003} (ES). Most of the distinction is due to the more complex, yet restricted, structure of ES. In Ghirardato (2002), DC is defined for all events with respect to the ex-ante preference. In ES, DC is defined only between one time and the next, and therefore at each point in time it applies to fewer events. That is, in ES the DM's information is represented by a filtration and dynamic consistency requires that whenever the DM ``anticipates'' preferring one act to another in each realizable contingency tomorrow, then she must also prefer it today. This form of dynamic consistency requires the set of priors to have a ``rectangular'' structure with respect the filtration. In \autoref{reversal}, a violation of the dynamic consistency from \citet{Epstein2003} is exhibited; under PB both $g\succsim_{\{y\}}f$ and $g \succsim_{\{r,b\}} f$, yet $f \succ_S g$.}   While dynamic consistency has normative appeal (see \citet{Ghirardato2002} for additional discussion), it is typically violated in models of ambiguity. Further, MEU preferences cannot satisfy dynamic consistency unless restrictions are placed on the conditioning events and the structure of the set of priors (e.g., rectangularity).  Indeed, the preference pattern in \autoref{reversal}  violates dynamic consistency as $f=fAg \succ g$ and $g \succ_A f$.

\begin{example}\label{reversal-two} Recall the setup in \autoref{reversal} and consider the constant act $x=(2.5, 2.5, 2.5)$. Then $gAx=(10,0,2.5)\sim x$.  How should she rank $g$ and $x$ after learning $A$ has occurred? As $x$ is a constant act, the revelation that $A$ has occurred cannot increase the value of $x$. But the revelation that $A$ has occurred may improve the evaluation of $g$ if $g$ is ``good in $A$.''  The ex-ante preference $gAx \succsim x$ reveals that $g$ is ``good in $A$,'' and so she should prefer $g$ to $x$: $g\succsim_A x$.\footnote{It is even plausible that she would have a strict preference $g \succ_A x$, but it should never be the case that she strictly prefers $x$ to $g$ after $A$.}%
\end{example}

In line with the intuition of \autoref{reversal-two}, I introduce the following weakening of dynamic consistency.

\begin{axiom}[Weak Planning Consistency]\label{WUDC} For each $A \in \Sigma$ and for all $f,x \in \mathscr{F},$ \[fAx \succsim x \implies f \succsim_A x.\]
\end{axiom}

Behaviorally, we can think of Kate expressing that she is unwilling to commit to a choice between two uncertain objects. However,  she feels confident in her ability to compare an act against a sure-thing. Thus under certain conditions she would be willing to commit \emph{ex-ante} to choose $f$ over $x$ if $A$ were realized.

\nameref{WUDC} weakens dynamic consistency in two ways.\footnote{\nameref{WUDC} also weakens the conditional certainty equivalent consistency condition introduced by \citet{pires2002} to characterize FB.} First, it only retains the forward-looking planning condition (i.e., part (i)). Second, it only applies to plans which involve comparison with a ``sure'' comparison. It is worth noting that $f$ may be strictly preferred to $x$ after $A$ if the agent uses the observation of $A$ to ``rule out'' implausible priors.  Indeed, following \autoref{reversal-two} it is easy to calculate that under PB with $\rho(A)=\frac12$, $gAx \sim x$ and $g \succ_A x$.

Additionally, it is straightforward to see that the backward-looking condition of dynamic consistency (i.e., part (b)) fails. For instance, consider $z=(3,3,3)$. Then following the calculations in \autoref{reversal}, $g \sim_A z$ and $z \succ gAz$.  The logic behind this is quite natural. Even if Kate anticipates that she would prefer $g$ to $z$ if $A$ occurred, she does not want to commit to that before any information is revealed. This unwillingness to commit might stem from a lack of confidence in certainty probability distributions. From the ex-ante view, she has no idea which event will occur, and so given her initial set of beliefs she prefers the certain payoff of $z$ to the uncertain payoff of $gAz$.

\subsection{Objective randomizations}

The final axiom necessary to characterize PB links the unconditional and conditional preferences via a consistency condition on her desire to reduce subjective uncertainty to objective uncertainty. Therefore, a bit more notation is in order. 

\begin{definition}Let $\alpha \in \Delta(S)$.  Then for any $f \in \F$, denote by $f^{\alpha}$ the constant act that yields the mixture outcome of $f$ according to $\alpha$:% on $X$ that returns $f(s)$ with probability $\alpha(s)$ for each state $s \in S$; 
$$f^{\alpha}:=\left(\sum_{s}\alpha(s)f(s)\right)\mathbf{1}_{S}.$$
\end{definition}

 Similar notions of reducing subjective uncertainty to objective uncertainty via mixing have been used to distinguish multi-utility and multi-belief representations \citep{Ok2012} and study preferences which admit Allais and Ellsberg behavior \citep{dean2017hedge}. To connect to conditional preferences, it is necessary to define conditional objective mixtures.  For any objective randomization $\alpha \in \Delta(S)$, and any event $A\in \Sigma$, let $\alpha(A) = \sum_{s \in A}\alpha(s)$.  This is the objective probability given to $A$ under the objective randomization.  Given any objective randomization and event $A$, we can derive the ``updated'' randomization that only gives weight to states in $A$. For every $\alpha \in \Delta(S)$ and any $A \subset S$ such that $\alpha(A)>0$, let $\alpha_A$ denote the objective randomization such that $\alpha_A(s)=\frac{\alpha(s)}{\alpha(A)}$ for $s \in A$ and $0$ otherwise. 

% \begin{definition} Let $\alpha \in \Delta(S)$. Say that the agent prefers objective uncertainty at $\alpha$ if $f^{\alpha} \succsim f$ for any $f \in \F$. 
%\end{definition}

The following proposition shows that for every unambiguously $\succsim$-nonnull $A$, there exists a randomization $\alpha_A \in \Delta(S)$ such that (i) $\alpha_A(A)=1$ and (ii) the agent always prefers to reduce an act to a constant according to $\alpha_A$. This is a simple consequence of the fact that $\succsim_A$ admits an MEU representation. However, willingness to eliminate subjective uncertainty with a conditional objective randomizations can be linked to the willingness to eliminate subjective uncertainty with objective randomizations under the ex-ante preference.  The following proposition shows that a consistency condition on the desire to reduce subjective uncertainty to objective uncertainty before and after information provides an alternative characterization of FB.

%\begin{proposition}\label{FBprop} Suppose $\{\succsim_A\}_{A\in\Sigma}$ satisfies \nameref{CMEU} and \nameref{CON}. Then there exists $\alpha_A \in \Delta(S)$, such that $\alpha_A(s)=0$ for $s \in S\setminus A$, and the agent prefers objective uncertainty at $\alpha_A$. %$f^{\alpha_A} \succsim_A f$ for all $f \in \F$. 
%Further, for each strongly $\succsim$-nonnull $A$ the following are equivalent: 
%\begin{itemize}
%\item[(i)] The collection $\{\succsim_A\}_{A\in\Sigma}$ admits a partial bayesian representation with $\rho(A)=0$ 
%\item[(ii)] For every $\alpha \in \Delta(S)$, both of the following hold:
%	\begin{itemize}
%	\item[(a)] If the agent prefers objective uncertainty at $\alpha$, then the agent prefers objective uncertainty at $\alpha_A$.%If $f^{\alpha} \succsim f$ for all $f \in \F$, then  $f^{\alpha_A} \succsim_A f$ for all $f \in \F$
%	\item[(b)] If the agent prefers objective uncertainty at $\alpha_A$, then the agent prefers objective uncertainty at $\alpha$ for some $\alpha \in [\alpha_A]$. %If $f^{\alpha_A} \succsim_A f$ for all $f \in \F$, then $f^{\alpha} \succsim f$ for all $f \in \F$ for some $\alpha \in [\alpha_A]$. 
%	\end{itemize}
%\item[(iii)] For every $f,x \in \F$, $fAx \sim x \LRA f \sim_A x.$
%\end{itemize}
%\end{proposition}
\begin{proposition}\label{FBprop} Suppose $\{\succsim_A\}_{A\in\Sigma}$ satisfies \nameref{CMEU} and \nameref{CON}. Then for each $A \in \Sigma$, there exists $\alpha_A \in \Delta(S)$, such that $\alpha(s)=0$ for $s \in S\setminus A$, and  $f^{\alpha_A} \succsim_A f$ for all $f \in \F$. 
Further, for every  $A \in \Sigma$ that is unambiguously $\succsim$-nonnull, the following are equivalent: 
\begin{itemize}
\item[(i)] $\succsim_A$ admits an MEU representation with FB updating $(u,\mathscr{C}_A^{FB})$.
\item[(ii)] For every $f \in \F$ and $\alpha \in \Delta(S)$, both of the following hold:
	\begin{itemize}
	\item[(a)] If $f^{\alpha} \succsim f$ for all $f \in \F$, then  $f^{\alpha_A} \succsim_A f$ for all $f \in \F$
	\item[(b)] If $f^{\alpha_A} \succsim_A f$ for all $f \in \F$, then $f^{\alpha} \succsim f$ for all $f \in \F$ for some $\alpha \in [\alpha_A]$. 
	\end{itemize}
\item[(iii)] For every $f,x \in \F$, $fAx \sim x \text{ if and only if } f \sim_A x$.
\end{itemize}
\end{proposition}

Conceptually, this proposition shows that under FB there is a tight link between an agent's willingness to convert subjective uncertainty to objective uncertainty before and after information. The equivalence between (i) and (iii) is established in \cite{pires2002}, and shows that FB imposes a consistency condition on conditional certainty equivalents. However, under PB with $\rho(A)>0$ for some event, the agent may violate (iii) (and also dynamic consistency) because she makes an inference about her priors, given the information observed, and discards priors that fail her likelihood test. 

Before introducing the final axiom, consider the following scenario. Before Kate is allowed to choose an act, she is offered the choice to implement a randomization device to reduce any act to a lottery. She is presented with two devices, $\alpha$ and $\alpha'$, and she states that she is happy to reduce subjective uncertainty, for any possible act, with either of these devices.  After she receives partial information, $A$, she is asked again how she feels about the devices. For simplicity, suppose $\alpha(A)>\alpha'(A)$.  

Imagine that Kate were to state that she was still happy to use $\alpha'$, but was unwilling still use $\alpha$. Kate thinks twice and asks her friend, Alice for advice. Alice says ``that is peculiar, because your willingness to use $\alpha$, for any act $f$, indicated that you were happy with the odds $\alpha$ provided in $A$. Because $\alpha(A)>\alpha'(A)$, the odds $\alpha$ provided in $A$ had a bigger impact on the lottery that you would receive. Now that $A$ has happened, you should also be happy with $\alpha$.''  Put differently, if Kate is still confident enough in a particular expert to continue soliciting their advice, she ought to continue to heed the advice from any expert that was ``more accurate.'' 

If Kate finds Alice's reasoning persuasive, she will satisfy the following axiom, \nameref{URC}, that requires precisely this form of monotonicity. 

%\begin{axiom}[Monotone Reduction Consistency]\label{URC} Suppose $\alpha,\alpha' \in \Delta(S)$ are such that the agent prefers objective uncertainty at $\alpha$ and $\alpha'$ and $\alpha(A) \ge \alpha'(A)$. %$f^{\alpha} \succsim f$ and $f^{\alpha'} \succsim f$ for all $f \in \F$ and any $A \in \Sigma$, 
%Then whenever the agent prefers objective uncertainty at $\alpha'_A$, the agent also prefers objective uncertainty at $\alpha_A$. 
%
%\end{axiom}
\begin{axiom}[Monotone Reduction Consistency]\label{URC} Suppose $\alpha,\alpha' \in \Delta(S)$ are such that $f^{\alpha} \succsim f$ and $f^{\alpha'} \succsim f$ for all $f \in \F$ and $\alpha(A) \ge \alpha'(A)$, then
\[f^{\alpha'_A} \succsim_A f \text{ for all } f \in \F  \implies f^{\alpha_A} \succsim_A f \text{ for all } f \in \F.\]

%$\alpha(A) \ge \alpha'(A)$ and $f^{\alpha'_A} \succsim_A f$ for all $f\in\mathscr{F}$, then $f^{\alpha_A} \succsim_A f$ for all $f\in\mathscr{F}.$
\end{axiom}

\nameref{URC} requires that when (i) objective randomizations can be ordered by their weight on $A$, and (ii) the agent still wishes to eliminate ambiguity through $\alpha'$, which placed a lower weight on $A$ than did $\alpha$,  after $A$ has occurred, then she must also prefer to eliminate ambiguity via $\alpha$. 

This axiom is a weakening of the reduction conditions in \autoref{FBprop}. To see this, note that by \autoref{FBprop}, FB requires both $f^{\alpha_A}\succsim_A f$ and $f^{\alpha'_A}\succsim_A f$. Hence, FB requires that the agent always prefers to reduce subjective uncertainty according to both $\alpha_A$ and $\alpha'_A$, no matter how they are related.

\begin{theorem}\label{PBU} Suppose the collection of preferences $\{\succsim_A\}_{A\in\Sigma}$ satisfies \nameref{CMEU} and \nameref{CON}. Then the following are equivalent:  
\begin{itemize}
\item[(i)] For every unambiguously $\succsim$-nonnull $A \in \Sigma$, the preference $\succsim_A$ satisfies \nameref{WUDC} and \nameref{URC} 
\item[(ii)] The collection of preferences $\{\succsim_A\}_{A\in\Sigma}$ admits a partial bayesian updating representation.

\end{itemize}
\end{theorem}

The uniqueness properties of the representation are summarized in the following proposition. 

\begin{proposition}\label{uniqueness} Suppose that $(u, \mathscr{C},\rho)$ and $(u', \mathscr{C}',\rho')$ both represent the same preferences. Then $u=au' + b$ for some $a,b \in \R, a >0$, $\mathscr{C}=\mathscr{C}'$. Further, for each $A \in \Sigma$ that is unambiguously $\succsim$-nonnull, exactly one of the following cases holds:
\begin{itemize}

\item[(i)] there exist $f,x \in \F$, such that $fAx \sim x \text{ and } f  \succ_A x$, and then \[\rho(A)=\rho'(A).\]

\item[(ii)] for all $f,x \in \F$, $fAx \sim x \text{ if and only if } f  \sim_A x$, and then \[\rho(A), \rho'(A) \in  \left[0, \rho^*(A)\right],\]
for \[\rho^*(A)=\min_{\pi_A \in BU(\mathscr{C},A)}\left(\frac{\max_{\pi \in [\pi_A]\cap\mathscr{C}}\pi(A)}{\max_{\mu \in \mathscr{C}}\mu(A)}\right).\]
\end{itemize}
\end{proposition}

\begin{remark}
Uniqueness of $\rho(A)$ follows from the uniqueness of $\mathscr{C}$, and it can only be point identified when $\succsim_A$ violates the conditional certainty equivalent consistency condition of \cite{pires2002}. Such a violation happens when $\mathscr{C}_A^{FB} \neq \mathscr{C}_A^{ML} $ and $\rho(A) > \rho^*(A)$.  Observe that if $\mathscr{C}_A^{FB} = \mathscr{C}_A^{ML}$, then for every $\pi \in \mathscr{C}$ there must be a $\pi' \in \mathscr{C}$ such that $\pi_A=\pi_A'$ and $\pi'(A)=\max_{\mu \in \mathscr{C}}\mu(A)$. Consequently, for each $\pi_A \in \mathscr{C}_A$, $\max_{\pi \in [\pi_A]\cap\mathscr{C}}\pi(A)=\max_{\mu \in \mathscr{C}}\mu(A)$ and thus $\rho^*(A)=1$. It is obvious that whenever $\mathscr{C}_A^{FB} \neq \mathscr{C}_A^{ML}$, it must be the case that $\rho^*(A)<1$.

%The potential for stronger identification arises when there is an event $A$ and priors $\mu,\mu' \in \mathscr{C}$ such that $\mu(A)=0$ and $\mu'(A)>0$.  As can be seen in \autoref{uniqueness}, if we allow $\min\pi(A)=0$, then we no longer have an issue with identification for any $\rho(A)>0$ and FB precisely coincides with $\rho(A)=0$. However, it is common for axiomatic models of updating to only consider unambiguously non-null events, so in this case it is not clear how to extend updating to the case of ``possibly null events." \footnote{For instance, \citet{Ghirardato2008} and \citet{Hanany2007} restrict attention to unambiguously non-null events. The assumption of strict monotonicity in \nameref{CMEU} rules out zero-probability states, so that every state (hence every event) is unambiguously non-null: $\mu(A)>0$ for all $\mu \in \mathscr{C}$.  Strict monotonicity could be relaxed and the axioms I have provided could be imposed on the collection of unambiguously non-null events without substantial changes. However, this would not improve uniqueness.}

\end{remark}

Note that non-uniqueness of $\rho(A)$ arises only when $ \rho(A) \in \left[0,\rho^*(A)\right]$, in which case PB is behaviorally indistinguishable from FB. Intuitively, we should think about these cases as simply being cases of a zero inference threshold and force $\rho(A)$ to its minimal value.  Accordingly, a form of uniqueness can be obtained by defining minimal representations. 

\begin{definition}Say that $(u,\mathscr{C},\rho)$ is a minimal PB representation of $\{\succsim_A\}_{A\in\Sigma}$ if for any $(u', \mathscr{C}',\rho')$ that also represents preferences, $\rho(A) \leq \rho'(A)$. \end{definition}

A minimal representation requires $\rho(A)=0$ whenever $\succsim_A$ satisfies the conditional certainty equivalent consistency condition of \cite{pires2002} ($fAx \sim x \text{ if and only if } f  \sim_A x)$. Clearly any minimal representation is unique.

\section{Properties of partial bayesian updating}\label{discussion}

In this section, I establish a method for comparing changes in belief sets between individuals and show that inference has a sharp relation to ambiguity attitude. I provide a formal comparison of PB updating with RML, showing that RML violates the essential monotonicity condition of PB. I then show that PB results in a primacy effect; hence the order of information arrival matters. Lastly, I characterize PB with a constant inference threshold. 

%\subsection{Unambiguous Preference}
%
%Consider the derived \emph{unambiguously preferred} relation of \cite*{Ghirardato2004}:
%\begin{definition}  Say that f is \emph{unambiguously preferred} to g, denoted $f \succsim^* g$, if and only if for any act h and any $\lambda \in [0,1]$, $\lambda f + (1-\lambda)h \succsim \lambda g + (1-\lambda)h$. \end{definition}
%
%It is known that this relation admits a 
%
%However, such examples all involve violations in a ``single direction.'' That is, note that $g$ and $h$ are not comparable according to the unambiguous preference ex-ante. After making inferences, the set of priors shrinks and so some acts which were previously incomparable under $\succsim^*$ are now comparable under $\succsim^*_A$. However, consider two acts such that $fAg \succsim^* g$. Then $f$ is "unambiguously better in $A$" and therefore no mater how much inference the agent makes, she must prefer $f$ to $g$ if $A$ is realized. Hence I impose the following weakening of dynamic consistency. 
%
%\begin{axiom}[Weak Unambiguous Dynamic Consistency]For each $A \in \Sigma$ and for all $f, g \in \mathscr{F},$ \[fAg \succsim^* g \implies f \succsim^*_A g.\]
%\end{axiom}
%

\subsection{Comparative inference and ambiguity attitude}

This section develops a comparative statics notion to compare agents' inference thresholds.  That is, what choice behavior is consistent with a greater $\rho$?  This comparison is partially confounded when agents differ in their prior beliefs.  
However, whenever $\mathscr{C}_{2,S} = \mathscr{C}_{1, S}$, it follows that $\mathscr{C}^{\rho_1}_{1, A} \subseteq  \mathscr{C}^{\rho_2}_{2,A}$ if $\rho_1(A) \geq \rho_2(A)$. When both agents reveal inference at $A$ (e.g., $\rho_1(A) \geq \rho_2(A)$ are uniquely identified), then the converse also holds:  $\rho_1(A) \geq \rho_2(A)$ if $\mathscr{C}^{\rho_1}_{1, A} \subseteq  \mathscr{C}^{\rho_2}_{2,A}$.  Consequently, I will assume that unconditional preferences are identical. 

Intuitively, if agent one makes more inferences than two, agent one is less concerned with the residual uncertainty than agent two. This intuition may be formalized by defining a comparative notion of less uncertainty averse.  
 
\begin{definition}Say that $\succsim_1$ is \textbf{less uncertainty averse} than $\succsim_2$ if for all $f,x \in \F$,
$$x \succsim_1 (\succ_1) f \RA x \succsim_2 (\succ_2) f.$$
Further, whenever $\succsim_1$ and $\succsim_2$ are ordinally equivalent on constant acts (e.g., admit the same risk preference), then $\succsim_1$ is \textbf{less ambiguity averse} than $\succsim_2$. 
\end{definition}

In essence, this states that whenever agent one prefers a sure-thing, so must agent two. Hence agent two is more averse to uncertainty, and agent one is less so. This is the definition provided in \citet{GhirardatoMarinacci2002}. When risk preferences are the same between the agents, then this notion also characterizes relative ambiguity aversion.\footnote{\citet{GhirardatoMarinacci2002} require a notion of ``cardinal symmetry,'' which is satisfied in the current setting when $u_1$ is a positive affine transformation of $u_2$, i.e.,  $x \succsim_1 y \LRA x \succsim_2 y$.}

\begin{theorem}\label{WTIC}Consider two agents that admit a minimal partial bayesian representations $(u, \mathscr{C}, \rho_1)$ and $(u, \mathscr{C}, \rho_2)$.  Then the following are equivalent,
\begin{itemize}
\item[(i)]  $\succsim_{1,A}$ is less uncertainty averse than $\succsim_{2,A}$,
\item[(ii)] $\rho_1(A) \geq \rho_2(A)$.
%\item[(iii)]  $\succsim^*_{1,A}$ is more complete than $\succsim^*_{2,A}$ for every $A \in \Sigma$,
\end{itemize}
\end{theorem}

One implication of \autoref{WTIC} is that PB may lead to a reduction in perceived ambiguity and that increased inference leads to less ambiguity. In the context of financial markets, reduced perception of ambiguity may lead to increased stock market participation after news.\footnote{This was also remarked  by \citet{Epstein2008ambiguity}.} Further, those who infer more (i.e., have a higher $\rho(A)$) will participate more. 

\begin{remark}This result also suggests potential for an alternative axiomatization of PB.  Since the agent's posterior set under PB is  (weakly) smaller than FB, the agent's conditional reference, $\succsim_A$, will be (weakly) less ambiguity averse.  Thus a condition that regulates \emph{reduced ambiguity aversion} may exist that could replace \nameref{URC}. I leave progress in this direction to others.  
\end{remark}

\begin{corollary} Let $\succsim_{FB(A)}$, $\succsim_{ML(A)}$, and $\succsim_{A}$ denote preferences corresponding to the FB, ML, and PB updating rules conditional on $A \in \Sigma\setminus S$, respectively. Then $\mathscr{C}_A^{ML} \subseteq \mathscr{C}_A^{\rho} \subseteq \mathscr{C}_A^{FB}$ and hence
\begin{itemize}
\item[(i)] $\succsim_{ML(A)}$ is less ambiguity averse than $\succsim_{A}$
\item[(ii)] $\succsim_{A}$ is less ambiguity averse than $\succsim_{FB(A)}$ 
\end{itemize}
\end{corollary}

\subsection{Comparison to RML}\label{rml}

To better understand PB, it is instructive to compare it to the RML of \cite{Cheng2021}. Under RML, the agent first forms a selection by contracting her set of priors and then updating each of the selected priors. Since this selection is not based on likelihood, it may violate \nameref{URC}. Formally, let $\mathscr{C}^*(A)=\{p \in \mathscr{C} \mid p(A)=\max_{p' \in \mathscr{C}} p'(A)\}$, denote the subset of $\mathscr{C}$ which consists of those priors that assign maximal probability to $A$. Then let $\mathscr{C}_{\gamma_A}(A)=(1-\gamma_A)\mathscr{C} + \gamma_A \mathscr{C}^*(A)$ denote the contraction of $\mathscr{C}$ towards $ \mathscr{C}^*(A)$. Then a (contingent) relative maximum likelihood representation is defined as follows: 

\begin{definition}The collection $\{\succsim_A\}_{A \in \Sigma}$ is represented by contingent relative maximum likelihood (cRML) updating if $\succsim$ has MEU representation $(u, \mathscr{C})$ and for all unambiguously $\succsim$-nonnull $A \in \Sigma$ there exists $\gamma_A \in [0,1]$ such that $\succsim_A$ has MEU representation $(u, \mathscr{C}_A)$ with \[\mathscr{C}_A = \{ \pi_A \mid \pi \in  \mathscr{C}_{\gamma_A}(A)\}, \]
and $\mathscr{C}_{\gamma_A}(A)=(1-\gamma_A)\mathscr{C} + \gamma_A \mathscr{C}^*(A)$. If $\gamma_A=\gamma$ for all $A$, it is referred to as the RML. 
\end{definition} 

While both PB and cRML result in a sets of conditional beliefs that are ``between'' FB and ML, they differ in how this selection is made. In particular, under PB a belief is ``retained'' based on likelihood alone and therefore its retention rule respects monotonicity. On the other hand, cRML may violate this monotonicity, as illustrated in the following example. 
%Finally, this set is updated, prior-by-prior, according to Bayes' rule. %so that after $E$ the DM is MEU with respect to $$\Pi_{C_{\alpha_E}(E)}:=\{q | q = \frac{p}{p(E)} \text{ for some } p \in C_{\alpha_E}(E)\}.$$

\begin{figure}[h]
\centering
\includegraphics[height=7cm]{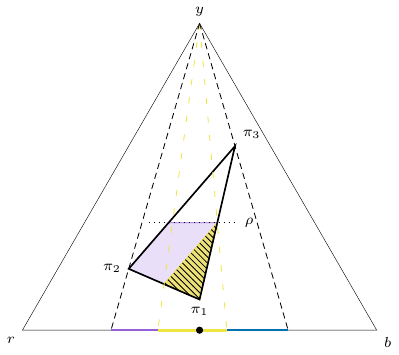}
\caption{PB and RML posterior beliefs after $A=\{r,b\}$.}\label{RMLfig}
\end{figure}

\begin{example} Recall the setup of \autoref{reversal}. In \autoref{RMLfig}, the larger shaded region (purple) represents the selection under PB (for $\rho\approx 0.72$), while the smaller, patterned  triangle (yellow) represents selection under the cRML model with $\gamma_A=1/2$. The corresponding posterior sets are illustrated also: PB results in the region between the left-most dashed line and the right, loosely dashed line (union of the purple and yellow regions), while RML is given by the region between the two (interior) loosely dashed lines (the yellow region). Note that under RML there are priors $\pi, \pi' \in \mathscr{C}$ such that (i) $\pi(A) > \pi'(A)$, and (ii) $\pi'$ is updated but $\pi$ is not updated after $A$. 

For concreteness, note that RML excludes $\pi_2=(6/10, 2/10, 2/10)$ but retains $\pi=(11/40, 15/40, 14/40)$ (which the midpoint between $\pi_1$ and $\pi_3$). This violates \nameref{URC}, since $\pi_2(\{r,b\})=32/40 > 26/40 = \pi(\{r,b\})$. In fact, the entire shaded region between the far left dashed line and the left, loosely dashed line (yellow) consists of priors that the agent rejects under cRML and lead to violations of \nameref{URC}. 
 %For these priors, the Bayesian update of $\pi_A \in \Pi_{C_{1/2}(E)}$, where $\pi_2$ is the Bayesian update of $\pi_2$, while $\pi_1^A=(3/4,1/4)$ is not.

%In fact, there is a large set of such probabilities, \autoref{RMLfig}

\end{example}

\subsection{Updating and the primacy effect}

One implication of Bayesian updating with a single prior is that posterior beliefs only depend on the final information. Thus beliefs are independent of the order in which information arrives.  This property of \emph{informational path independence} can be defined for a set of beliefs, independently of the specific updating rule, and is formalized below. 

\begin{definition}Beliefs satisfy \textbf{informational path independence} (IPI) if for any $B \subset A \in \Sigma$, $$\left(\mathscr{C}_A\right)_B=\mathscr{C}_{ B},$$
where $\left(\mathscr{C}_A\right)_B$ denotes the update of $\mathscr{C}_A$ given $B$. 

\end{definition}

When the order of information does matter, it may be that earlier or later information has a greater effect. If the earlier information matters more, then there is a \textbf{primacy effect}; if the later information matters more, then there is a \textbf{recency effect}.  In general, the PB rule does not satisfy IPI. Since posterior sets are formed by applying Bayes rule to a selection of priors, it will be the case that the PB exhibits a primacy effect, but not a recency effect, as illustrated by the following example.

\begin{example}\label{IPIexample}
Let $S= \{s_1, s_2, s_3,s_ 4\}, A=\{s_1, s_2, s_3\},$ and $B=\{s_1,s_2\}$.  Suppose 
 \[\mathscr{C} =\left\{ \left(\frac14, \frac14-\beta, \frac14 + 2\beta, \frac14-\beta \right) \mid \beta \in [0,\frac18],\right\}\] and, for simplicity, $\rho(A)= \rho(B) = 1$.  Through direct calculation
 \[\mathscr{C}_A^{\rho} = \{(\frac{2}{7},\frac{1}{7},\frac{4}{7},0)  \}, \mathscr{C}_B^{\rho} =\{(\frac{1}{2},\frac{1}{2},0,0)\}, \text{ but } 
(\mathscr{C}_A^{\rho})_B^{\rho} = \{(\frac{2}{3},\frac{1}{3},0,0)\}.\]
Hence IPI is violated at $A$ and $B$. This arises because of the ``opposing inferences'' made after $A$ and $B$, respectively. That is, a ML decision maker concludes that $\beta$ is large after $A$, but concludes that $\beta$ is small after $B$. If the decision maker sequentially revises her beliefs, her initial inference is ``carried forward,'' resulting in a primacy effect. 

Relaxing $\rho(A)=1$ illustrates how this earlier inference influences $(\mathscr{C}_A^{\rho})_B^{\rho}$. 
\begin{itemize}
\item When $\rho(A) \in [0,\frac67]$, no priors are excluded from $\mathscr{C}_A^{\rho}$, and so IPI will not be not violated. 
\item When $\rho(A) \in (\frac67,1]$, a prior is selected and updated after $A$ only if $\beta \ge \rho(A)\frac{7}{8} - \frac{3}{4}$. Since $\rho(B)=1$, after $B$ only those priors that minimize $\beta$ are selected. Thus \[(\mathscr{C}_A^{\rho})_B^{\rho} = \left\{\left(\frac{2}{10-7\rho(A)},\frac{8-7\rho(A)}{10-7\rho(A)},0,0\right)\right\}.\]
\end{itemize}
Since beliefs after $B$ depend critically on which priors where selected after $A$ (via $\rho(A)$), a primacy effect is exhibited.
\end{example}

The violation in \autoref{IPIexample} occurs because in the prior for which event $A$ is maximized, the probability ratio of $s_1$ and $s_2$ is different than in the prior for which event $B$ is maximized.\footnote{This feature was anticipated by \citet{Epstein2007}, who assume a similar functional form to PB to study long-run learning. Because of this feature, they assume that in each period the agent forms a ``fresh'' posterior set from the initial set of priors $\mathscr{C}$, rather than sequentially apply the updating rule.} More generally, whether PB violates informational path independence may depend on the structure of $\mathscr{C}$ and the exact magnitude of $\rho$. For instance, for rectangular $\mathscr{C}$ (relative to some information structure),  FB and ML coincide and therefore informational path independence is satisfied for every PB representation. In the absence of rectangularity, informational path independence will hold for FB, as stated in the following proposition.

\begin{proposition}\label{IPIprop}Suppose the collection of preferences $\{\succsim_A\}_{A\in\Sigma}$ admits a minimal PB representation. If $\rho(A)=0$ for all unambiguously $\succsim$-nonnull $A \in \Sigma$, then the agent satisfies informational path independence for every pair of events. 
\end{proposition}

As a complementary result, I show that PB updating will ``typically'' violate IPI and exhibit a primacy effect. To do so, I establish a condition on the set of priors for when there is some event and inference thresholds for which IPI is violated.  Recalling the intuition of \autoref{IPIexample}, a violation of IPI requires that the threshold at $A$ is sufficiently large so that it excludes priors that would have passed the test for inclusion in $B$.  This is formalized below. 

\begin{proposition}\label{IPIprop2} Suppose for some events, $B \subset A$, $A\neq B$, $|B| \ge 2$, there exists $\underline{\rho} \in [0,1]$, such that for some prior $\pi \in \mathscr{C}$ and $s,s' \in B$,  $\frac{\pi(s)}{\pi(s')} \neq \frac{\mu(s)}{\mu(s')}$ for all $\mu\in\{ \mu \in \mathscr{C} \mid \mu(A)\ge \underline{\rho} \max_{\hat{\mu}\in \mathscr{C}} \hat{\mu}(A) \}$. Then there exists $\overline{\rho} \in [0,1]$ that that a PB updater violates IPI whenever $\rho(B) \leq \overline{\rho}$  and $\rho(A)\ge\underline{\rho}$.
\end{proposition}

Note that this is only possible when there is a violation of rectangularity (\citet{Epstein2003}). Therefore another way to interpret this result is that when FB and ML result in distinct sets of posteriors (i.e., rectangularity is violated), there exist thresholds for which PB will violate IPI. Consequently, the only way to ensure compliance with IPI is to utilize FB updating.

\subsection{Event-independent inference}

To characterize an event-independent inference threshold, I require a stronger version of \nameref{URC}.  The requisite strengthening must impose a form of monotonicity across events, rather than just within an event.  In order to introduce this strengthening, I require a bit more notation.  

\begin{definition}For every $A \in \Sigma$, say that $\alpha \in \Delta(S)$ is \textbf{$A$-maximal} if $f^{\alpha} \succsim f$ for all $f \in \F$ and for any other $\alpha'  \in [\alpha_A]$ such that $f^{\alpha'} \succsim f$ for all $f \in \F$,  $\alpha(A) \geq \alpha'(A)$. 
\end{definition}

That is, given some distribution over $A$, we consider the objective randomization that is ``identical within $A$'' and puts maximal likelihood on $A$. A necessary, though not sufficient, condition for being $A$-maximal is being on the boundary of the set of priors. It should be noted that this definition does not preclude the existence of two $A$-maximal randomizations such that $\alpha(A) > \alpha'(A)$.  This is possible so long as they do not map to the same conditional randomization: $\alpha_A \neq \alpha'_A$.  Put another way, there must be states $s,s' \in A$ such that $\frac{\alpha(s)}{\alpha(s')}\neq \frac{\alpha'(s)}{\alpha'(s')}$. Before stating the final axiom, I require one more definition. 

\begin{definition} For every $A \in \Sigma$ for some $x,y \in X$, with $x \succ y$, define $\overline{m}_A \in [0,1]$ by the indifference relation $\overline{m}_A y + (1- \overline{m}_A)x \sim yAx$. 
\end{definition}

The value $\overline{m}_A$ captures the agent's subjective, maximal probability of $A$ and is independent of $x$ and $y$.  With these two concepts established, I now introduce the additional axiom.

\begin{axiom}[Monotone Likelihood Reduction Consistency]\label{ORC} For all $A,B \in \Sigma$, consider any $\alpha,\alpha' \in \Delta(S)$ such that they are $A$-maximal and $B$-maximal, respectively, and $\frac{\alpha(A)}{\overline{m}_A} \geq \frac{\alpha'(B)}{\overline{m}_B}$. Then 
\[f^{\alpha'_B} \succsim_B f \text{ for all } f\in\mathscr{F} \implies f^{\alpha_A} \succsim_A f \text{ for all }  f\in\mathscr{F}.\]
\end{axiom}

\nameref{ORC} states that if an agent prefers to reduce an act to a lottery when given $\alpha'$ ($\alpha'_B$), before (after) learning some event $B$ and also prefers to reduce acts to lotteries via $\alpha$, then whenever $\alpha$ puts objectively higher weight on $A$ than $\alpha'$ does on $B$ (when normalized by $\overline{m}$)  the agent also desires to reduce subjective uncertain with $\alpha_A$ after $A$.  Note that when $A=B$, this reduces to \nameref{URC}. Hence this axiom provides both within-event restrictions on updating behavior (if $A=B$) and between-event consistency (when $A\neq B$).

\begin{theorem}\label{cPBU} The following are equivalent:  
\begin{itemize}
\item[(i)] The collection of preferences $\{\succsim_A\}_{A\in\Sigma}$ satisfies \nameref{CMEU}, \nameref{CON}, \nameref{WUDC}, and \nameref{ORC} for every unambiguously $\succsim$-nonnull $A \in \Sigma$.
\item[(ii)] The collection of preferences $\{\succsim_A\}_{A\in\Sigma}$ admits a partial bayesian updating representation with a constant $\rho$.

\end{itemize}
\end{theorem}

\section{Conclusion}

This paper axiomatizes a model of updating a set of priors that generalizes both FB and ML.  Upon receiving information, the agent makes an inference about her priors and applies Bayes' rule to the sufficiently plausible priors. Since PB provides a one-parameter generalization of both FB and ML, it could reasonably find use in applications where these updating rules have been utilized (\citet{Beauchene2019}, \citet{Bose2014}, and \citet{Kellner2018}). 

As is common in models of ambiguity, the PB rule (and thus also FB and ML) may result in violations of dynamic consistency.  This is an unavoidable cost if we want an updating procedure that may be applied to any set of priors and any event. Because PB is behaviorally characterized by weakening dynamic consistency to allow for inference, PB provides a rationale for these violations of dynamic consistency. Indeed, dynamic consistency is violated because the agent uses the information to distinguish between priors, rather than treating all priors as equal.  Thus under ambiguity, inference about priors necessitates dynamic preference reversals that a decision maker might find defensible. 

By characterizing updating behavior through a consistency condition on the agent's desire to reduce subjective uncertainty to objective uncertainty, the approach I have taken may be useful to understand ``inference'' in other models of ambiguity.  However, I leave this for future work. 
  
\appendix

\section{Proofs}

First, I will introduce some notation and state without proof some basic results which are used in all following proofs.  
\begin{itemize}
%\item For any $\Pi \subseteq \Delta(S)$ and $A$ such that $ \pi (A) >0$ for all $\pi \in \Pi$, $BU(\Pi, A)=\{ \pi_A \mid \pi \in \Pi\}$. 
\item For any $u:X \ra \R$, let $\mathscr{U}=u(X) \subset \R$.  
\item For any $a \in \mathscr{U}^{|S|} \subset \R^{|S|}$, it is clear that there exists some $f \in \F$ such that $(u\circ f )(s)=a(s)$.
%\item[(i)] The collection of preferences $\{\succsim_A\}_{A\in\Sigma}$ satisfies \nameref{CMEU} if and only if for each $A \in \Sigma$, there exists a closed, convex set of priors $\mathscr{C}_A$ and a non-constant affine function $u_A:X \rightarrow \mathbb{R}$ so that \[f \succsim_A g \LRA \min_{\pi \in \mathscr{C}_A} \sum_{s \in S}u_A(f(s))\pi(s) \geq \min_{\pi \in \mathscr{C}_A} \sum_{s \in S}u_A(g(s))\pi(s).\]  
%\item[(ii)] If the collection additionally satisfies ordinal preference consistency, we can suppose without loss that for all $A$, $u_A = u_{S}$.  
%\item[(iii)] If $A$ is strongly $\succsim$-nonnull, then $\pi(A)>0$ for all $\pi \in \mathscr{C}$.
\end{itemize}

Next, I will state a basic lemma that will be used several times throughout the paper. 

\begin{lemma}\label{sht} Suppose $\succsim_A$ admits an MEU representation $(u, \mathscr{C}_A)$. For any objective randomization $\alpha \in \Delta(S)$, if $f^{\alpha_A} \succsim_A f$ for all $f \in \F$, then there is some $\pi \in \mathscr{C}_A$ such that $\pi = \alpha_A$.  
\end{lemma}

\begin{proof}
To see this, suppose $\alpha_A \neq \pi$ for all $\pi \in \mathscr{C}_A$. Then $\alpha_A \notin\mathscr{C}_A$. Since $\mathscr{C}_A$ is a closed and convex subset of $\R^{|S|}$, by the separating hyperplane theorem there is some $a \in \R^{|S|}$ such that $\sum_{s \in A}a(s)\alpha(s) < \min_{\pi \in \mathscr{C}_A}\sum_{s \in A}a(s)\pi(s)$. By certainty independence we can without loss assume that $a \in \mathscr{U}^{|S|}$, hence there exists $f \in \F$ so that $u\circ f = a$. Hence $\sum_{s \in A}u(f(s))\alpha(s) = u(\sum_{s \in A}f(s)\alpha(s)) < \min_{\pi \in \mathscr{C}_A}\sum_{s \in A}u(f(s))\pi(s) \LRA f^{\alpha_A} \prec_A f$.  This is a contradiction, hence here is some $\pi \in \mathscr{C}_A$ such that $\pi = \alpha_A$. 
\end{proof}

\subsection{Proof of \autoref{FBprop}}

 \begin{proof}
\noindent{\textbf{Existence of $\alpha_A$:}} By \nameref{CMEU} there is a closed, convex set $\mathscr{C}_A \subset \Delta(S)$ such that $(u, \mathscr{C}_A)$ represents $\succsim_A$. By \nameref{CON}, for any $x,y \in \F, xAy \sim_A x$. Hence for every $\mu \in \mathscr{C}_A$, $\mu(S\setminus A)=0$. Choose any $\mu \in \mathscr{C}_A$ and let $\alpha_A(s)=\mu(s)$ for $s \in A$. Then \[u(f^{\alpha_A})= u((\sum_{s}\alpha_A(s)f(s))\mathbf{1}_{S}) =\sum_{s}\alpha_A(s)u(f(s)) \geq \min_{\mu \in \mathscr{C}_A} \sum_{s}\mu(s)u(f(s)),\] where the last equality follows from Independence (\nameref{CMEU}) and the inequality follows from the represetnation and the fact that $\alpha_A \in \mathscr{C}_A$. Since $f$ was arbitrary,  $\alpha_A$ satisfies $f^{\alpha_A} \succsim_A f$ for all $f \in \F$. Further, this establishes that if $\pi \in \mathscr{C}_A$, then for any randomization such that $\alpha_A(s)=\pi(s)$ for all $s \in A$, $f^{\alpha_A} \succsim_A f$ for all $f \in \F.$

\noindent{\textbf{Characterization of FB:}} That $(i) \LRA (iii)$ can be found in \cite{pires2002}. To see $(ii) \RA (i)$: consider any $\alpha$ such that $f^{\alpha} \succsim f$ for all $f \in \F$. It follows from \autoref{sht} that this holds if and only if there is some $\mu \in \mathscr{C}$ such that $\alpha=\mu$. By similar reasoning, $f^{\alpha_A} \succsim f$ for all $f \in \F$ ensures $\mu_A \in \mathscr{C}_A$, and so $BU(\mathscr{C},A) \subset \mathscr{C}$. To see containment in the other direction, suppose $\pi_A \in \mathscr{C}_A$. Thus $f^{\pi_A} \succsim_A f$ for all $f \in \F$. By (ii.b), there is a $\mu \in \mathscr{C}\cap [\pi_A]$, but then $\pi_A=\mu_A \in \mathscr{C}_A$. Thus $BU(\mathscr{C},A) = \mathscr{C}_A$ and $(i)$ holds. The proof of $(i)\RA (ii)$ is similar. 

\end{proof}

\subsection{Proof of \autoref{PBU}}\label{thrmproof}

\begin{proof}It is obvious that $(ii) \RA (i)$, hence I will only prove that $(i) \RA (ii)$.  We restrict attention to $A$ that are unambiguously $\succsim$-nonnull. 

{\bf Step 1:} First, I establish that $\mathscr{C}_A \subseteq BU(\mathscr{C}, A)$. If $\succsim_A$ satisfies $fAx \sim x \LRA f \sim_A x$, then the representation holds for $\rho(A)=0$. % If we relax strict monotonicity and allow for $\pi(A)=0$ for some $\pi \in \mathscr{C}$, then $\rho=0$ if and only if $\succsim^*$ satisfies DC} 
In what follows, suppose $\succsim_A$ violates this property for some $A$. Then let $\mathcal{K}_{\succsim}=\{A \in \Sigma \mid fAx \sim x \mbox{ and } f \succ_A x\}$ denote the set of events at which preferences reveal inference at $A$. Further, it is clear that \nameref{WUDC} implies that for any $A$, $\mathscr{C}_A \subset BU(\mathscr{C}, A)$. To see this, suppose $\mu \in \mathscr{C}_A \setminus BU(\mathscr{C}, A)$.   By \autoref{sht} there is some $a \in  \R^{|S|}$ such that $\sum_{s \in A}a(s)\mu(s) < \min_{\pi \in \mathscr{C}_A}\sum_{s \in A}a(s)\pi(s)$, hence there exists $f \in \F$ so that \[\sum_{s \in A}u(f(s))\mu(s) < \min_{\pi \in BU(\mathscr{C}, A)}\sum_{s \in A}u(f(s))\pi(s).\] Let $x$ solve $u(x)= \min_{\pi \in BU(\mathscr{C}, A)}\sum_{s \in A}u(f(s))\pi(s)$, and thus by construction $fAx\sim x$ but $x \succ_A f$. Hence $\mathscr{C}_A \subset BU(\mathscr{C}, A)$. 

%Hence for every $\mu \in \mathscr{C}_{A}$ there is some $\pi \in \mathscr{C}$ such that $\pi_A = \mu$

%{\bf Step 2:} For any objective randomization $\alpha \in \Delta(S)$, if $f^{\alpha_A} \succsim_A f$ for all $f \in \F$, then there is some $\pi \in \mathscr{C}_A$ such that $\pi = \alpha_A$.  To see this, suppose $\alpha_A \neq \pi$ for all $\pi \in \mathscr{C}_A$.  Then since $\mathscr{C}_A$ is closed and convex, by the separating hyperplane theorem %there is some $a \in  \R^{|S|}$ such that $\sum_{s \in A}a(s)\alpha(s) < \min_{\pi \in \mathscr{C}_A}\sum_{s \in A}a(s)\pi(s)$.  By certainty independence we can without loss assume that $a \in K^{|S|}$, hence 
%there exists $f \in \F$ so that $\sum_{s \in A}u(f(s))\alpha(s) = u(\sum_{s \in A}f(s)\alpha(s)) < \min_{\pi \in \mathscr{C}_A}\sum_{s \in A}u(f(s))\pi(s) \LRA f^{\alpha_A} \prec_A f$.  Further, it is clear that if $\pi \in \mathscr{C}_A$, then for any randomization such that $\alpha_A(s)=\pi(s)$ for all $s \in A$, $f^{\alpha_A} \succsim_A f$ for all $f \in \F.$

{\bf Step 2:} Next, I will establish that $\mathscr{C}_A = \{\pi_A \mid \pi \in \mathscr{C} \text{ and } \pi(A) \geq \rho(A)\max_{\mu \in \mathscr{C}}\mu(A)\}$ for some $\rho(A) \in [0,1]$. Intuitively, I find this number by employing the fact that $A \in \mathcal{K}_{\succsim}$, and thus $\mathscr{C}_A $ is a strict subset of $ BU(\mathscr{C}, A)$. Indeed, this means that there is some $ \beta \in \Delta(S)$ such that $\beta_A \in BU(\mathscr{C}, A)$, and $\beta_A \notin \mathscr{C}_A $. When we ``project back'' to $\mathscr{C}$, we identify a ``slice''  $[\beta_A]\cap \mathscr{C}$ of $\mathscr{C}$ that was discarded. I look at these slices to find the right $\rho(A)$. 

Formally, for every $A \in \mathcal{K}_{\succsim}$ let $\varepsilon_{A}=\sup\{\pi(A)| \pi \in \mathscr{C} \mbox{ and } \pi_A \notin \mathscr{C}_A\}$. This number exists because $A \in \mathcal{K}_{\succsim}$. By construction, if $\pi \in \mathscr{C}$ and $\pi(A)\geq \varepsilon_{A}$, then $\pi_A \in \mathscr{C}_{A}.$    Further, since $A\in \mathcal{K}_{\succsim}$, it follows that \[0<\min_{\pi \in \mathscr{C}}\pi(A) < \varepsilon_A \leq \max_{\pi \in \mathscr{C}}\pi(A).\]

Note this does not rule out some $\pi$ so that $\pi(A) < \varepsilon_A$ and $\pi_A \in \mathscr{C}$.  However, if this is the case then it must be that there is some other $\pi'$ so that $\pi(A) \geq \varepsilon_A$ and $\pi'_A=\pi_A$. To see this, suppose to the contrary. Then there is some $\mu \in \mathscr{C}_A$ such that for every $\pi \in \mathscr{C}$ satisfying $\pi_A=\mu$, $\pi(A)<\varepsilon_A$. By definition of $\varepsilon_A$, and the fact that $\mathscr{C}$ is closed, there must be some $\theta \in \mathscr{C}$ such that $\theta_A \notin \mathscr{C}_A$ and $\theta(A) > \max_{\pi \in [\pi_A]\cap\mathscr{C}} \pi(A)$, or else the supremum property is violated. Then for any $\alpha \in [\pi_A]\cap\mathscr{C}$,  $\theta(A) \geq \alpha(A)$, $f^{\alpha}\succsim f$, $f^{\theta}\succsim f$, $f^{\alpha_{A}} \succsim_{A}f,$ where the last follows from the hypothesis that $\pi_A=\mu  \in \mathscr{C}_A$. Then by \nameref{URC}, it must be that $f^{\theta_A} \succsim_{A} f$, a contradiction of $\theta_A  \notin \mathscr{C}_A$. Hence $\mu \in \mathscr{C}_A$ only if there is some $\pi \in \mathscr{C}$ so that $\pi(A) \geq \varepsilon_A$ and $\pi_A=\mu$.

Now let $\rho(A)$ be defined by $\rho(A)=\frac{\varepsilon_A}{\max_{\pi \in \mathscr{C}}\pi(A)}$ if $A \in \mathcal{K}_{\succsim}$, and $\rho(A)=0$ if $A \in \Sigma\setminus \mathcal{K}_{\succsim}$. Thus \[\mathscr{C}_A^{\rho}:=  \{\pi_A \mid \pi(A) \geq \rho(A) \max_{\mu \in \mathscr{C}}\mu(A)  \text{ and }  \pi \in \mathscr{C}\}=\mathscr{C}_A.\]

Note that this defines a minimal representation $(u, \mathscr{C}, \rho)$. 
\end{proof}

\subsection{Proof of \autoref{uniqueness}}

\begin{proof}Suppose $(u,\mathscr{C}, \rho)$ and $(u',\mathscr{C}', \rho')$ are two PB representations of the same collection of preferences. Uniqueness of $u$ and $\mathscr{C}$ are standard due to the MEU representation of $\succsim$. Now consider some unambiguously $\succsim$-nonnull event $A$. 

{\bf Case (ii):}  Suppose first that for all $f,x \in \F$, $fAx \sim x \text{ if and only if } f  \sim_A x$. Then necessarily $\mathscr{C}_A^{\rho} = BU(\mathscr{C},A)=BU(\mathscr{C}',A)=\mathscr{C}_A^{\rho'}$.  Consequently, for each $\pi_A \in BU(\mathscr{C},A)$, $[\pi_A]\cap \mathscr{C} \neq \varnothing$ and thus for both $\rho,\rho'$, 
\[\rho(A)\max_{\mu \in \mathscr{C}}\mu(A) \leq \max_{\pi \in [\pi_A]\cap \mathscr{C}} \pi(A),\]
and
\[\rho'(A)\max_{\mu \in \mathscr{C}}\mu(A) \leq \max_{\pi \in [\pi_A]\cap \mathscr{C}} \pi(A).\]

Since each of these inequalities must hold for every $\pi_A \in BU(\mathscr{C},A)$, it follows immediately that \[\rho(A), \rho'(A) \leq \rho^*(A):= \min_{\pi_A \in BU(\mathscr{C},A)}\left(\frac{\max_{\pi \in [\pi_A]\cap\mathscr{C}}\pi(A)}{\max_{\mu \in \mathscr{C}}\mu(A)}\right).\]

Each term of $\rho^*(A)$ is well-defined because all constraint sets are compact. 

{\bf Case (i):}  Now, suppose there exist $f,x \in \F$, such that $fAx \sim x \text{ and } f  \succ_A x$. The fact that $\rho(A)= \rho'(A)$ then follows immediately from fact that $\mathscr{C}_A^{\rho'}= \mathscr{C}_A^{\rho} $ and that each of $\rho(A), \rho'(A)$ are uniquely determined by $\mathscr{C}$ and $\mathscr{C}_A$ (which follows from the construction in Step 2 of Theorem 1).

%\neq \mathscr{C}_A^{FB}$. Since $\mathscr{C}_A^{\rho} \subset \mathscr{C}_A^{FB}$, 

%Following from $(ii)$, it must be that $\rho(A), \rho'(A) > \rho^*(A)$. Suppose $\rho(A) \neq \rho'(A)$, and thus without loss we may assume that $\rho(A) > \rho'(A)$.  

%Since $ \mathscr{C}_A^{ML} \subseteq \mathscr{C}_A^{\rho} \subseteq \mathscr{C}_A^{FB}$, it also immediately follows that $ \mathscr{C}_A^{ML} \neq  \mathscr{C}_A^{FB}$.

\end{proof}

\subsection{Proof of \autoref{WTIC}}

\begin{proof} Since both agents satisfy the representation and $\succsim_1 = \succsim_2$,  we can conclude that $(u_1,\mathscr{C}_1)=(u_2,\mathscr{C}_2)=(u,\mathscr{C})$.  We can without loss restrict attention to $A$ such that $fAx \sim x$ and $f \succ_{i,A} x$ for both $i$. If there is no such $A$, then since we supposed a minimal PB representation, $\rho_1(A)=\rho_2(A)=0$ always and $\succsim_{1,A}=\succsim_{2,A}$. If it is only violated by one of the agents, the proof is nearly identical. 

{\bf Step 1 $(ii)\RA(i):$} First, it is trivial that if $\rho_1(A) \geq \rho_2(A)$, then $\mathscr{C}_A^{\rho_1} \subseteq \mathscr{C}_A^{\rho_2}$.  Consequently, suppose $x  \succsim_{1,A} f$. Then 
$$u(x) \geq \min_{\pi \in \mathscr{C}_A^{\rho_1}}\sum_{s \in A}u(f(s))\pi(s) \geq \min_{\pi \in \mathscr{C}_A^{\rho_2}}\sum_{s \in A}u(f(s))\pi(s),$$
where the second inequality follows from $\mathscr{C}_A^{\rho_1} \subseteq \mathscr{C}_A^{\rho_2}$. Hence $x\succsim_{2,A}f$ and $\succsim_{1,A}$ is less uncertainty averse than $\succsim_{2,A}$.

{\bf Step 2 $(i)\RA(ii):$} Suppose $\succsim_{1,A}$ is less uncertainty averse than $\succsim_{2,A}$. Then by previous results it follows that $\mathscr{C}_A^{\rho_1} \subseteq \mathscr{C}_A^{\rho_2}$. It is straightforward from here that $\rho_1(A) \geq \rho_2(A)$.

\end{proof} 

\subsection{Proof of \autoref{IPIprop}}

%That (ii) and (iii) are equivalent is straightforward, given that we are assuming a minimal representation. 
When $\rho(A)=0$ for every $A$, the agent updates every belief. Let $B \subset A$ and suppose $\mu \in \mathscr{C}$. Then $\mu(B)>0 \implies \mu(A) > 0$ and $\mu_A(B)=\frac{\mu(B)}{\mu(A)}$.  Consequently, $\mu_B(s) = \frac{\mu(s)}{\mu(B)}= \frac{\mu(s)}{\mu(A)}\frac{\mu(A)}{\mu(B)} = \frac{\mu_A(s)}{\mu_A(B)}=(\mu_A)_B(s)$. Since $\mu$ was arbitrary, it follows that $\mathscr{C}_B= (\mathscr{C}_A)_B$. %To see the reverse, suppose (iii) is false. Then 

\subsection{Proof of \autoref{IPIprop2} }

\begin{proof}Let $B=\{s,s'\} \subset A$. By assumption $\pi(A) < \rho(A)\max_{\hat{\mu}\in \mathscr{C}} \hat{\mu}(A)$. Further, since $\frac{\pi(s)}{\pi(s')} \neq \frac{\mu(s)}{\mu(s')}$ for all $\mu$ for which $\mu(A)$ is sufficiently high, there is no $\mu_A \in \mathscr{C}_A$ satisfying $\frac{\pi(s)}{\pi(s')} = \frac{\mu_A(s)}{\mu_A(s')}$. Hence $\pi_A \notin \mathscr{C}_A$, and consequently $\pi_B \notin( \mathscr{C}_A)_B$. Then let $\overline{\rho}= \frac{\pi(B)}{\max_{\hat{\mu}\in \mathscr{C}} \hat{\mu}(B)}$. Then whenever $\rho(B) \leq \overline{\rho}$, it follows that  $\pi_B \in \mathscr{C}_B$. Thus IPI is violated at $A$ and $B$.
\end{proof}

\subsection{Proof of \autoref{cPBU}}

\begin{proof}

When $A=B$, \nameref{ORC} implies \nameref{URC}, hence for each $A$ there exists $\rho(A)$ such that $\mathscr{C}_A^{\rho}=\mathscr{C}_A$. 

{\bf Step 1:} We know that for each $A\in \mathcal{K}_{\succsim}$ there is an $\varepsilon_A$ so that $\pi(A) \geq \varepsilon_A$ implies $\pi_A \in \mathscr{C}_A$.  Consider any two $A,B \in \mathcal{K}_{\succsim}$.  Then we have both $$\min_{\pi \in \mathscr{C}}\pi(A) < \varepsilon_A \leq \overline{m}_A \text{ and } \min_{\pi \in \mathscr{C}}\pi(B) < \varepsilon_B \leq \overline{m}_B.$$

Now, let $\alpha$ be $A$-maximal and $\alpha'$ be $B$-maximal and $f^{\alpha'_{B}} \succsim_{B}f $ for every $f$. By \autoref{ORC}, if $\frac{\alpha(A)}{\overline{m}_A} \geq \frac{\alpha'(B)}{\overline{m}_B} \LRA \alpha(A) \geq \alpha'(B)\frac{\overline{m}_A}{\overline{m}_B}$, then there is some $\pi \in \mathscr{C}$ such that $\pi=\alpha$ and $\pi_A \in \mathscr{C}_A$, hence 
$$\varepsilon_A\geq\alpha'(B)\frac{\overline{m}_A}{\overline{m}_B} \geq \varepsilon_B\frac{\overline{m}_A}{\overline{m}_B}.$$
By symmetry we may conclude that $\frac{\varepsilon_A}{\overline{m}_A} = \frac{\varepsilon_B}{\overline{m}_B}$ for all $A,B \in \mathcal{K}_{\succsim}$. Hence we may define $\rho := \frac{\varepsilon_A}{\overline{m}_A},$ and it follows that $\rho(A)=\rho$ for every $A \in\mathcal{K}_{\succsim}$.

\begin{figure}[h]
\centering
\includegraphics[height=6cm]{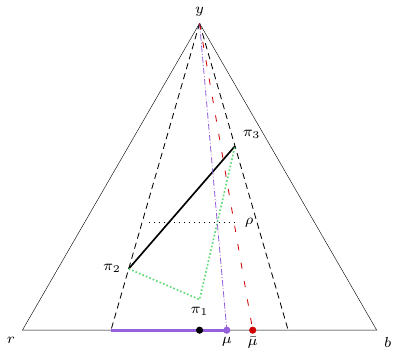}
\caption{The collection of $\{r,b\}$-maximal randomizations is given by the (green) dotted boundary of $\mathscr{C}$. The posterior set $\mathscr{C}_{\{r,b\}}^\rho$ is the (solid purple line) region at the base of the simplex between the left-most dashed line and the (purple) dash-dotted line. The intersection of the (purple) dash-dotted  line and $\mathscr{C}$ indicates all priors that result in the posterior $\mu \in \mathscr{C}_{\{r,b\}}^\rho$. The intersection of the (red) loosely dashed line and $\mathscr{C}$ indicates all priors that result in the posterior $\hat{\mu} \notin \mathscr{C}_{\{r,b\}}^\rho$.}\label{Proof-fig}
\end{figure}

{\bf Step 2:} Now the final step is to show that if $A \notin\mathcal{K}_{\succsim}$, then $\mathscr{C}_A=\mathscr{C}_A^\rho$.  Since $A \notin \mathcal{K}_{\succsim}$, it follows that $\mathscr{C}_A= BU(\mathscr{C},A)$, and clearly $\mathscr{C}_A^\rho \subset BU(\mathscr{C},A)$.  Suppose for contradiction that there is some $\mu \in  BU(\mathscr{C},A)$ such that $\mu \notin \mathscr{C}_A^\rho.$  Then let $\pi \in \mathscr{C}$ satisfy $\pi_A=\mu$. Next, let $\alpha$ be an $A$-maximal randomization so that $\alpha_A(s)=\mu(s)$ for all $s \in A$. It then follows from our hypothesis that $\alpha(A) < \rho \overline{m}_A$. Next, consider any $B \in \mathcal{K}_{\succsim}$. Since $B \in \mathcal{K}_{\succsim}$, there is some $g,x$ such that $gBx \sim x$ and $g \succ_B x$, hence $BU(\mathscr{C},B) \setminus \mathscr{C}_B \neq \emptyset$ and for all $\pi \in BU(\mathscr{C},B) \setminus \mathscr{C}_B$, it must be that $\pi(B) <  \rho \overline{m}_B$. If follows that there is some $B$-maximal $\hat{\alpha}$ so that $\hat{\alpha}(B) < \rho \overline{m}_B$. Since $\mathscr{C}$ is closed and convex, we can without loss suppose $\frac{\hat{\alpha}(B)}{\overline{m}_B} \geq \rho -\theta$ for any $\theta >0$. To see why, consider \autoref{Proof-fig}, and note that by taking $\bar{\mu}$ arbitrarily close to $\mu$, the (red) dashed line traces out a sequence of $\{r,b\}$-maximal distributions $\alpha$ such that $\alpha(\{r,b\})$ converge to $\rho\overline{m}_B$.  Consequently, we may suppose that $\frac{\hat{\alpha}(B)}{\overline{m}_B}> \frac{\alpha(A)}{\overline{m}_A}$. But, by assumption $f^{\alpha} \succsim f$ and $f^{\alpha_A} \succsim f$ for every $f \in \F$, and therefore by \nameref{ORC} we require $f^{\hat{\alpha}_B} \succsim f$ for every $f$, which implies that there is some $\hat{\mu} \in \mathscr{C}_B$ with $\hat{\alpha}_B (s)=\hat{\mu}(s)$ for $s \in B$. But, since $\hat{\alpha}$ is $B$-maximal and $B \in \mathcal{K}_{\succsim}$, this must mean that $\hat{\alpha}(B) \geq \rho \overline{m}_B$.  This contradicts our selection of $\hat{\alpha}(B)$, hence $\mathscr{C}_A=\mathscr{C}_A^\rho$. 

%assumption that there is some $\mu \in  BU(\mathscr{C},A)$ such that $\mu \notin \mathscr{C}_A^\rho.$ Hence $\mathscr{C}_A=\mathscr{C}_A^\rho$. 

%we can without loss take $\hat{\alpha}$ so that $|\hat{\alpha}(B)-\overline{m}_B| < \frac{\overline{m}_A - \alpha(A)}{\overline{m}_A}\overline{m}_B$. 

\end{proof}

\bibliographystyle{ecta}
%\bibliography{/Users/matthew/dropbox/references/references.bib} %iMac
\bibliography{/Users/matthewkovach/dropbox/References/references.bib} %MBP
%\bibliography{C:/Users/Matt/references}

\end{document}